\documentclass[journal]{IEEEtran}
\usepackage{cite,amssymb}
\usepackage[pdftex]{graphicx}
\usepackage{algorithm}
\usepackage[cmex10]{amsmath}
\usepackage{algorithm}
\usepackage{algpseudocode}
\usepackage{array,balance}
\usepackage{bm}

\ifCLASSOPTIONcompsoc
\usepackage[caption=false,font=normalsize,labelfont=sf,textfont=sf]{subfig}
\else
\usepackage[caption=false,font=footnotesize]{subfig}
\fi
\usepackage{url}
\usepackage{amsfonts}
\usepackage{exscale}
\usepackage{relsize}
\usepackage{multicol,color}
\newcommand{\mH}{\mathop{\rm H}}
\newcommand{\mT}{\mathop{\rm T}}
\begin{document}
\title{OPARC: Optimal and Precise Array Response Control Algorithm -- Part II: 
	Multi-points\\ and Applications}
\author{Xuejing Zhang,~\IEEEmembership{Student Member,~IEEE,}
	Zishu He,~\IEEEmembership{Member,~IEEE,}
	Xiang-Gen Xia,~\IEEEmembership{Fellow,~IEEE,}\\
	Bin Liao,~\IEEEmembership{Senior Member,~IEEE,}
	Xuepan Zhang, and
	Yue Yang,~\IEEEmembership{Student Member,~IEEE}
	\thanks{X. Zhang, Z. He and Y. Yang are with the University of Electronic Science and Technology of China, Chengdu 611731, China (e-mail: xjzhang7@163.com; zshe@uestc.edu.cn; yueyang@std.uestc.edu.cn).}
	\thanks{X. Zhang and X.-G. Xia are
		with the Department of Electrical and Computer Engineering,
	University of Delaware, Newark, DE 19716, USA (e-mail: xjzhang@udel.edu; xxia@ee.udel.edu).}
	\thanks{B. Liao is with College of Information Engineering, Shenzhen University, Shenzhen 518060, China (e-mail: binliao@szu.edu.cn).}
	\thanks{X. P. Zhang is with Qian Xuesen Lab of Space Technology, Beijing 100094, China (e-mail: zhangxuepan@qxslab.cn).}}
\markboth{}
{Shell \MakeLowercase{\textit{et al.}}: Bare Demo of IEEEtran.cls for Journals}
\maketitle


\begin{abstract}
In this paper, the optimal and precise array
response control (OPARC) algorithm proposed in Part I of this two paper series is
extended from single point to multi-points.
Two computationally attractive parameter determination approaches are provided
to maximize the array gain under certain constraints.
In addition, the applications of the multi-point OPARC algorithm to
array signal processing are studied.
It is applied to realize array pattern synthesis
(including the general array case and the large array case),
multi-constraint adaptive beamforming and quiescent pattern 
control, where an innovative concept of
normalized covariance matrix loading (NCL) is
proposed.
Finally, simulation results are presented to validate the superiority and effectiveness of the multi-point OPARC algorithm.
\end{abstract}

\IEEEpeerreviewmaketitle

\vspace*{-0.8\baselineskip}
\section{Introduction}
\IEEEPARstart{I}{n} the companion paper \cite{p1},
optimal and precise array response control
(OPARC) algorithm was proposed and analyzed.
OPARC provides a new mechanism to control array responses 
at a given set of angles, by simply assigning virtual interference
one-by-one.
The optimality (in the sense of array gain) of OPARC in each step is guaranteed.
Nevertheless, OPARC only controls one point per step and may be inefficient if multiple points
are needed to be precisely adjusted.
Moreover, how to use the OPARC algorithm in
practical cases (where real data commonly exists) remains.

This paper first extends the OPARC algorithm 
from single point response control per step
to multi-point response control per step. Note that a multi-point accurate array response control 
($ {\textrm M}{\textrm A}^2\textrm{RC} $) algorithm has been recently developed in \cite{ref201}.
Nevertheless, since it
is built on the basis of the accurate array response control ($ {\textrm A}^2\textrm{RC} $) algorithm \cite{snrf41},
the $ {\textrm M}{\textrm A}^2\textrm{RC} $ suffers from the similar drawbacks to $ {\textrm A}^2\textrm{RC} $, i.e.,
a solution is empirically adopted and hence 
a satisfactory performance cannot be always guaranteed
as analyzed in details in \cite{p1}.
In this paper, we first carry out a careful investigation on the change rule of the optimal beamformer when multiple virtual interferences
are simultaneously assigned.
Then, a generalized methodology of the weight vector update is observed
and utilized for the realization of the multi-point array response control.
Similar to the OPARC in \cite{p1}, we formulate
a constrained optimization problem such that the array response levels of multiple points can be optimally (in the sense of array gain) and precisely controlled.
Then, two different solvers, by either taking advantage of the OPARC algorithm or employing the recently developed consensus 
alternating direction method of multipliers (C-ADMM) approach in \cite{conADMM}, are provided to 
find an approximate solution
of the established optimization problem.
Note that since the OPARC in \cite{p1} only optimally controls the array response at one point in each step, it has a closed-form solution, while this is not the case for the multi-point OPARC in this paper. In other words, this paper does not cover \cite{p1}.
The differences between the proposed multi-point OPARC and 
$ {\textrm M}{\textrm A}^2\textrm{RC} $ are similar to those between OPARC and
$ {\textrm A}^2\textrm{RC} $ as described in [1] in details.
Meanwhile, for the proposed multi-point OPARC,
its applications to, such as, array pattern synthesis, multi-constraint adaptive beamforming
and quiescent pattern control, are also presented as detailed below.

{\it Application to Array Pattern Synthesis:}
Array pattern synthesis is a fundamental problem for radar, communication and remote sensing.
Most of the existing pattern synthesis approaches, for instance, 
the global optimization based methods in \cite{snrf07,snrf08,snrf09},
the convex programming (CP) methods in \cite{snrf13,snrf14,ref27}, and
the adaptive array theory based method in \cite{snrf12}, have no ability to precisely control the beampattern 
according to a given requirement.
In this paper, the above shortcoming is overcome by synthesizing desirable patterns with the proposed multi-point OPARC algorithm.
We start the synthesis procedure from the quiescent pattern, and iteratively adjust 
the responses of multiple angles to their desired levels. 
Simulation results show that it only requires a
few steps of iteration to complete the syntheses of well-shaped beampatterns.

In addition to the consideration for a general array,
large array pattern synthesis problem \cite{ref2007}, where the existing methods consume a large amount of computing resources or
even not work at all, is particularly discussed.
We will see that the large array pattern synthesis can be readily realized
with the multi-point OPARC algorithm, in a computationally attractive manner.

{\it Application to Multi-constraint Adaptive Beamforming:} Adaptive beamforming plays an important role in various application areas, since it enables us to receive a desired signal from a particular direction while it simultaneously blocks undesirable interferences.
Multi-constraint adaptive beamforming, i.e., designing an adaptive beamformer with 
several fixed directional constraints, is a common strategy to
improve the robustness of the adaptive beamformer, see \cite{lcmv,lcmv2,lcmv3} for
example.
The existing methods may cause distorted beampatterns, due to their
imperfections on model building or parameter optimization.
Based on the proposed multi-point OPARC algorithm, a new approach to
multi-constraint adaptive beamforming is presented in this paper.
We modify the traditional adaptive beamformer to make the
prescribed amplitude constraints satisfied by utilizing
the multi-point OPARC algorithm.
In the proposed algorithm, the total signal-to-interference-plus-noise ratio (SINR) (taking both 
real interferences and assigned virtual interferences into consideration) is maximized,
and the real unexpected components can be well rejected without leading to any undesirable pattern distortion.
Inspired by this, a new concept of normalized covariance matrix loading (NCL),
which can be regarded as a generalization of the conventional diagonal loading (DL) in \cite{ld1,ld2,ld3}, is developed.
Moreover, NCL is also exploited to realize quiescent pattern control as introduced next.

{\it Application to Quiescent Pattern Control:}
In brief, when an adaptive array operates in the presence of white noise only,
the resultant adaptive beamformer is referred to as the quiescent
weight vector, and the corresponding array response is termed as the
quiescent pattern.
As pointed out in \cite{con6}, having overall low sidelobes
is important to adaptive arrays and how to specify
a quiescent response pattern is worthwhile investigating.
Most of the existing quiescent pattern control methods \cite{con6,con200,con1} are established
on the foundation of the linearly constrained minimum variance (LCMV) framework,
where the unnecessary phase constraints of
array response are implicitly imposed.
In this paper, a simple yet effective quiescent pattern control algorithm
is proposed.
We synthesize a satisfactory deterministic pattern, i.e., the ultimate quiescent pattern,
by adopting the multi-point OPARC algorithm, and meanwhile, collect the resulting
virtual normalized covariance matrix (VCM) for later use.
Under the real data circumstance, the quiescent pattern control is completed by conducting a simple NCL operator to the existed VCM, and the
weight vector can be obtained accordingly.

This paper is organized as follows.
The proposed multi-point OPARC algorithm is presented in Section II.
The three applications of the multi-point OPARC are discussed in Section III. 
Representative experiments are carried out in Section IV and conclusions are drawn in Section V.

{\textit{Notations:}} The same as \cite{p1}, we use bold upper-case and lower-case letters to represent
matrices and vectors, respectively.
In particular, we use $ {\bf I} $ to denote the identity matrix.
$ j\triangleq\sqrt{-1} $.
$ (\cdot)^{\mT} $
and $ (\cdot)^{\mH} $ stand for the transpose and Hermitian
transpose, respectively.
$ |\cdot| $ denotes the absolute value and $ \|\cdot\|_2 $ denotes the $ l_2 $ norm.
We use $ ({\bf g})_i $ to
stand for the $ i $th element of vector $ {\bf g} $.
$ \Re(\cdot) $ and $ \Im(\cdot) $ denote the real
and imaginary parts, respectively. 
$ \oslash $ represents the element-wise division operator.
We use $ {\rm Diag}(\cdot) $ to stand for the diagonal matrix with the components of the input vector as the diagonal elements.
$ \mathbb{R} $ and $ \mathbb{C} $ denote the sets of all real and all complex numbers, respectively.
Finally, $ \cup $ denotes the set union and
$ {\rm card}(\cdot) $ returns the number of elements in a set.

\section{Multi-point OPARC Algorithm}
To present our multi-point OPARC algorithm, we first 
make a detailed analysis on the optimal weight vector.
\subsection{Multi-interference Optimal Beamformer}
Consider an array with $ N $ elements.
The same as \cite{p1}, the optimal weight vector:
\begin{align}\label{opt}
{\bf w}_{\rm opt}={\bf T}^{-1}_{n+i}{\bf a}(\theta_0)
\end{align}
maximizes both the output signal-to-interference-plus-noise ratio (SINR) and the array gain of an array system, 
where SINR and array gain are defined, respectively, as \cite{refbookada1}
\begin{align}\label{defGR}
{\rm SINR}\triangleq\dfrac{{\sigma}^2_s|{\bf w}^{\mH}{\bf a}(\theta_0)|^2}{{\bf w}^{\mH}{\bf R}_{n+i}{\bf w}},~~~
G\triangleq\dfrac{|{\bf w}^{\mH}{\bf a}(\theta_0)|^2}{{\bf w}^{\mH}{\bf T}_{n+i}{\bf w}}
\end{align}
where $ {\bf a}(\theta) $ stands for the array steering vector:
\begin{align}\label{atheta}
	{\bf a}(\theta)=[ g_1(\theta)e^{-j\omega\tau_1(\theta)},\cdots,g_N(\theta)e^{-j\omega\tau_N(\theta)}]^{\mT}
\end{align}	
where $ g_n(\theta) $ denotes the pattern of the $n$th element, $ \tau_n(\theta) $ is the time-delay between the $n$th element and 
	the reference point, $ n=1,\cdots,N $, $ \omega $ denotes the operating frequency.
In the above notations, $ \theta_0 $ is the beam axis,
${\bf R}_{n+i}$ denotes the $N\times N$ noise-plus-interference covariance matrix, 
$ {\bf T}_{n+i} $ stands for the normalized covariance matrix satisfying
\begin{align}\label{part201}
{\bf T}_{n+i}=\frac{{\bf R}_{n+i}}{\sigma_n^2}={\bf I}+\sum_{l=1}^{Q}{\beta}_l{\bf a}(\theta_l){\bf a}^{\mH}(\theta_l)
\end{align}
where $ {\beta}_l\triangleq{{\sigma}^2_l}/{{\sigma}^2_n} $ denotes the interference-to-noise ratio (INR), $ Q $ is the number of interferences,
$ {\bf a}(\theta_l) $ is the steering vector of the $ l $th interference,
$ {\sigma}^2_s $, $ {\sigma}^2_n $ and $ {\sigma}^2_l $ stand for the powers of
signal, noise and the $ l $th interference, respectively.
Note that $G$ in \eqref{defGR} represents the amplification factor of the input signal-to-noise ratio (SNR) $ {\sigma_s^2}/{\sigma_n^2} $, and the criterion of array gain maximization is adopted to achieve the optimal weight vector.

\newcounter{MYtempeqncnt}
\begin{figure*}[t]
	\normalsize
	\setcounter{MYtempeqncnt}{\value{equation}}
	\setcounter{equation}{14}
	\vspace*{-1.5\baselineskip}	
	\begin{align}\label{weightkk}
	{\bf w}_{k,\star}=
	{\bf w}_{k-1,{\star}}-{\bf T}^{-1}_{k-1}{\bf A}_k\left({\bf I}+{\bf \Sigma}_{k,\star}{\bf A}^{\mH}_k{\bf T}^{-1}_{k-1}{\bf A}_k\right)^{-1}
	{\bf \Sigma}_{k,\star}{\bf A}^{\mH}_k{\bf T}^{-1}_{k-1}{\bf a}(\theta_0)
	\end{align}
	\setcounter{equation}{\value{MYtempeqncnt}}
	\vspace*{-1.0\baselineskip}
	\hrulefill
\end{figure*}

From \eqref{opt}-\eqref{defGR}, one can see that the optimal weight vector 
$ {\bf w}_{\rm opt} $ depends on $ {{\bf R}_{n+i}} $ or $ {\bf T}_{n+i} $, which is 
normally data-dependent.
For this reason, 
$ {{\bf R}_{n+i}} $ or $ {\bf T}_{n+i} $ may not be available 
if we need to design a data-independent array response pattern $L(\theta,\theta_0)\triangleq{|{\bf w}^{\mH}{\bf a}(\theta)|^2}\big/{|{\bf w}^{\mH}{\bf a}(\theta_0)|^2}$
that satisfies some specific requirements.
In this case, for a given response design task,
the concept of virtual normalized noise-plus-interference covariance matrix (VCM)
was introduced in \cite{p1}.
Moreover, it was shown in \cite{p1} that
a VCM can be constructed by assigning suitable virtual interferences one-by-one.
In this paper, for a given response control task, we assign multiple virtual interferences
(instead of a single virtual interference)
at one step, and study how the optimal weight vector in \eqref{opt} changes.

We use induction to describe the problem.
Suppose that we have already assigned interferences for $ (k-1) $ times, the total
number of interferences is accumulated as $ Q_{k-1} $ and
$ {\bf T}_{k-1} $ denotes the total VCM
upto the $ (k-1) $th step.
The corresponding optimal weight vector at the $ (k-1) $th step is
given by
\begin{align}
{\bf w}_{k-1}={\bf T}^{-1}_{k-1}{\bf a}(\theta_0)
\end{align}
where the subscript $ (\cdot)_{\rm opt} $ has been omitted for notational simplicity.
Then, we carry out the $ k $th step by
assigning $ M_k $ interferences from directions $ \theta_{k,m} $ with INR to be 
$ {\beta}_{k,m} $, $ m=1,\cdots,M_k $,
where $ \theta_{k,m} $ are renamed from those $\theta_l$ in \eqref{part201}.
Then,
\begin{align}\label{Tk}
{\bf T}_k&={\bf T}_{k-1}
+\sum_{m=1}^{M_k}{\beta}_{k,m}{\bf a}(\theta_{k,m}){\bf a}^{\mH}(\theta_{k,m})\nonumber\\
&={\bf T}_{k-1}+{\bf A}_k{\bf \Sigma}_k{\bf A}^{\mH}_k
\end{align}
where
\begin{align}
{\bf A}_k&=[{\bf a}(\theta_{k,1}),\cdots,{\bf a}(\theta_{k,M_k})]\\
{\bf \Sigma}_k&={\rm Diag}([{\beta_{k,1},\cdots,{\beta}_{k,M_k}}])
\end{align}
and $ {\bf T}_k $ is the resulting VCM after implementing
the $ k $th step of the interference assigning. 
Clearly, if $ M_k=1 $, \eqref{Tk} degenerates to Eqn. (6) of \cite{p1},
and the related discussions return to our previous work in \cite{p1}.
To make the discussion meaningful,
the matrix $ {\bf A}_k $ in this paper is assumed to have a full column rank.
 
By applying the Generalized Woodbury Lemma \cite{refmat} to
\eqref{Tk}, we obtain that
\begin{align}\label{weitkk0}
{\bf T}^{-1}_{k}&=
{\bf T}^{-1}_{k-1}-\nonumber\\
&~~~{\bf T}^{-1}_{k-1}{\bf A}_k\left({\bf I}+{\bf \Sigma}_{k}{\bf A}^{\mH}_k{\bf T}^{-1}_{k-1}{\bf A}_k\right)^{-1}
{\bf \Sigma}_{k}{\bf A}^{\mH}_k{\bf T}^{-1}_{k-1}.
\end{align}
Accordingly, the obtained optimal weight vector satisfies
\begin{align}\label{weightk}
{\bf w}_{k}={\bf T}^{-1}_{k}{\bf a}(\theta_0)
={\bf w}_{k-1}+{\bf T}^{-1}_{k-1}{\bf A}_k{\bf h}_k
\end{align}
where $ {\bf h}_k\in\mathbb{C}^{M_k} $ is
\begin{align}\label{hh01}
{\bf h}_k=
-\left({\bf I}+{\bf \Sigma}_k{\bf A}^{\mH}_k{\bf T}^{-1}_{k-1}{\bf A}_k\right)^{-1}
{\bf \Sigma}_k{\bf A}^{\mH}_k{\bf T}^{-1}_{k-1}{\bf a}(\theta_0).
\end{align}
As shown in \eqref{weightk}, the current optimal weight $ {\bf w}_k $ is obtained
by making a modification to the previous weight $ {\bf w}_{k-1} $.

Recalling the adaptive array theory, the weight $ {\bf w}_{k} $
performs optimally in maximizing
the array gain $ G_k $ defined as
\begin{align}\label{part02}
G_k\triangleq{|{\bf w}^{\mH}_k{\bf a}(\theta_0)|^2}/{|{\bf w}^{\mH}_k{\bf T}_k{\bf w}_k|}
\end{align}
although the response levels
at $ \theta_{k,m} $, $ m=1,\cdots,M_k $, may not reach their expected values.
To precisely adjust the array responses of $ \theta_{k,m} $ to their
desired levels $ \rho_{k,m} $,
the INRs $ \beta_{k,m} $, $ m=1,\cdots,M_k $, or equivalently the 
diagonal matrix $ {\bf \Sigma}_k $, should be carefully selected.
In the meantime, the array
gain $ G_k $ in \eqref{part02} should be maximized.
Note also that $ {\bf h}_k $ in \eqref{hh01} acts as a mapping
of $ {{\bf \Sigma}_k} $, and we can express $ {\bf \Sigma}_k $ by $ {\bf h}_k $
as
\begin{align}\label{qua1197}
{\bf \Sigma}_k={\rm Diag}\left(-{\bf h}_{k}\oslash
\left({\bf A}^{\mH}_k{\bf T}^{-1}_{k-1}\left(
{\bf a}(\theta_0)+{\bf A}_k{\bf h}_{k}
\right)\right)\right).
\end{align}

From \eqref{hh01} and \eqref{qua1197}, one can see that
$ {\bf \Sigma}_k $ and $ {\bf h}_k $ are
one-one mapping. 
Therefore, the multi-point optimal and precise array response control (OPARC) can
be realized by either finding a suitable $ {\bf \Sigma}_k $ 
or selecting an appropriate $ {\bf h}_k $.

\subsection{Multi-point OPARC Problem Formulation}
Let us first formulate the multi-point
OPARC by optimizing $ {{\bf \Sigma}_k} $ as:
\begin{subequations}\label{qu101}	
	\begin{align}
	\max_{{\bf \Sigma}_k}&~~~G_k=
	{|{\bf w}^{\mH}_k{\bf a}(\theta_0)|^2}/{|{\bf w}^{\mH}_k{\bf T}_k{\bf w}_k|}\\
	{\rm subject~to}&~~~L(\theta_{k,m},{\theta_0})={\rho}_{k,m},~m=1,\cdots,M_k\\
\label{hhhhh}	&~~~{\bf w}_k={\bf w}_{k-1,\star}+{\bf T}^{-1}_{k-1}{\bf A}_k{\bf h}_{k}
	\end{align}
\end{subequations}
where 
$ {\bf w}_{k-1,\star} $ is the resultant weight vector of the $ (k-1) $th step
(we use the star symbol to indicate it as the ultimate selection of $ {\bf w}_{k-1} $),
the vector $ {\bf h}_{k} $ is given by \eqref{hh01}.
Once the optimal $ {\bf \Sigma}_{k,\star} $ has been obtained, we can express the
ultimate weight vector $ {\bf w}_{k,\star} $ as \eqref{weightkk} on the
top of this page.
To find the solution of problem \eqref{qu101}, an iterative method is 
first provided below.

\subsection{Iterative Approach}
The OPARC algorithm, developed in the companion paper \cite{p1}, is able to
optimally and precisely adjust one-point response level
at a time. Thus, we may apply it to
the $ M_k $-point OPARC problem \eqref{qu101} as follows.
For a fixed $ k>0 $, we apply the OPARC algorithm for $ M_k $ steps.
In the $ m $th step, OPARC is to realize $ L(\theta_{k,m},\theta_0)={\rho_{k,m}} $,
$ m=1,\cdots,M_k $.
Unfortunately, OPARC brings inevitable pattern variations on the previous controlled
angles as we have discussed in \cite{p1}.
More specifically, the response levels of $ \theta_{k,i} $, $ i=1,\cdots,m-1 $,
vary after accurately controlling the response level
of $ \theta_{k,m} $ to its desired level $ \rho_{k,m} $, $ 2\leq m\leq M_k $.
To reduce the undesirable pattern variations on the pre-adjusted angles, we propose 
to iteratively apply the $ M_k $-point OPARC for a number
	of times, until a certain termination criterion is met.
A temporary variable
	$ {\bf \Xi}={\bf T}_{k-1} $ and $ {{\bf \Sigma}_k}={\bf 0} $
are taken as the initializations in the first iteration. Then, in
each iteration, 
an $ M_k $-step OPARC is carried out. More specifically,
in the $ m $th step, we adjust the response level of $ \theta_{k,m} $ to be $ \rho_{k,m} $,
by calculating the INR of the newly assigned virtual interference
at $ \theta_{k,m} $, denoted as $ \beta_{k,m,\star} $,
$ m=1,\cdots,M_k $,
from Eqn. (38) of \cite{p1},
and then update the associated VCM as
$ {\bf \Xi}={\bf \Xi}+
{\beta}_{k,m,\star}{\bf a}(\theta_{k,m}){\bf a}^{\mH}(\theta_{k,m}) $.
Once an iteration, i.e., an $ M_k $-step OPARC, is completed, $ \beta_{k,m,\star} $
is added to the $ m $th diagonal element of $ {{\bf \Sigma}_k} $,
and then we set the resulting $ {\bf \Xi} $ as the initial VCM
	in the next iteration.
Note that $ {\bf T}_{0}={\bf I} $.

Naturally, whether the response levels of the adjusted angles $ \theta_{k,m} $, $ m=1,\cdots,M_k $, are close enough
to their desired levels can be a
criterion to terminate the iteration of OPARC.
However, this strategy needs to calculate
all the intermediate weight vectors that may be computationally inefficient.
To improve the computational efficiency, we propose 
to terminate the iteration of OPARC by examining whether
the magnitudes of INRs of the newly assigned virtual interferences approximate enough to zero,
since there is no need to assign virtual interferences
if their values are small enough.

Finally, we summarize the above iterative solver
of problem \eqref{qu101} in Algorithm \ref{cod1ed},
where $ {\beta}_{\epsilon} $ stands for a small tolerance parameter.
Note that
$ {\beta}_{k,m,\star} $ in Algorithm \ref{cod1ed} is calculated with 
Eqn. (38) of \cite{p1}.
In addition, we can express the ultimate $ {\bf \Sigma}_{k,\star} $ as
\addtocounter{equation}{1}
\begin{align}\label{INR1}
{\bf \Sigma}_{k,\star}={\rm Diag}([\bar{\beta}_{k,1,\star},\cdots,\bar{\beta}_{k,M_k,\star}])
\end{align}
where $ \bar{\beta}_{k,m,\star} $ represents the total
INR of the virtual interference assigned at $ \theta_{k,m} $ in the $ k $th step,
and
equals to the summation of all $ {\beta}_{k,m,\star} $'s 
of different iterations
for a fixed $ m=1,\cdots,M_k $.
As discussed earlier, once the optimal $ {\bf \Sigma}_{k,\star} $ has been obtained,
	we can use $ {\bf \Sigma}_{k,\star} $ to obtain
the VCM $ {\bf T}_k $ by Eqn. \eqref{Tk}, update $ {\bf h}_k $ in
\eqref{hh01} and \eqref{hhhhh}, and
calculate $ {\bf w}_{k,\star} $ by Eqn. \eqref{weightkk}.
It shall be noted that an inverse of 
normalized covariance matrix is indispensable in determining 
$ \beta_{k,m,\star} $'s by Eqn. (38) of \cite{p1}.
This may lead to a high cost in memory or/and computation especially
for a large array, although it may not need a large number of iterations.

\begin{algorithm}[!t]
	\caption{Iterative Approach to Problem \eqref{qu101}}\label{cod1ed}
	\begin{algorithmic}[1]
		\State give $ {\bf a}(\theta_0) $, 
		$ \theta_{k,m} $, $ {\rho_{k,m}} $, $ m=1,\cdots,M_k $, and $ {\bf A}_k $, set $ {\beta}_{\epsilon}>0 $,
		$ {\beta}_{\rm MAX}>{\beta}_{\epsilon} $, $ {\bf \Xi}={\bf T}_{k-1} $,
		$ {\bf \Sigma}_k={\bf 0} $
		\While{$ {\beta}_{\rm MAX}>{\beta}_{\epsilon} $}
		\For{$ m=1,\cdots,M_k $}
		\State calculate $ {\beta}_{k,m,\star} $ from Eqn. (38) of \cite{p1}, by setting
		$ ~~~~~~~~\!~~L(\theta_{k,m},\theta_0)={\rho}_{k,m} $ 
		\State update VCM $ {\bf \Xi}=
		{\bf \Xi}+
		{\beta}_{k,m,\star}{\bf a}(\theta_{k,m}){\bf a}^{\mH}(\theta_{k,m}) $
		\EndFor		
		\State update $ {\bf \Sigma}_k $ as
		$ {\bf \Sigma}_k={\bf \Sigma}_k+{\rm Diag}([{\beta_{k,1,\star},\cdots,{\beta}_{k,M_k,\star}}]) $
		\State obtain $ \beta_{\rm MAX}=\max\limits_{1\leq m\leq M_k}|{\beta}_{k,m,\star}| $
		\EndWhile
		\State obtain $ {\bf \Sigma}_{k,\star}={\bf \Sigma}_k $
	\end{algorithmic}
\end{algorithm}

\subsection{C-ADMM Approach}
We next propose another approach 
to solve problem \eqref{qu101}.
We first reformulate the original problem \eqref{qu101} as a 
quadratically constrained quadratic program (QCQP) problem.
Then,
the recently developed consensus-ADMM (C-ADMM) \cite{conADMM} approach is employed to 
find
its solution.

\subsubsection{Problem Reformulation}
Since $ {\bf h}_k $ is a one-one mapping of $ {\bf \Sigma}_k $, we can formulate
the multi-point OPARC, i.e., problem \eqref{qu101}, by finding $ {\bf h}_k $ as
\begin{subequations}\label{qu111}	
	\begin{align}
	\max_{{\bf h}_{k}\in\mathbb{C}^{M_k}}&~~~G_k=
	{|{\bf w}^{\mH}_k{\bf a}(\theta_0)|^2}/{|{\bf w}^{\mH}_k{\bf T}_k{\bf w}_k|}\\
\label{cons1}{\rm subject~to}&~~~L(\theta_{k,m},{\theta_0})={\rho}_{k,m},~m=1,\cdots,M_k\\
\label{cons}&~~~{\bf w}_k={\bf w}_{k-1,\star}+{\bf T}^{-1}_{k-1}{\bf A}_k{\bf h}_{k}.
	\end{align}
\end{subequations}
We
substitute the constraint \eqref{cons} into $ G_k $ and obtain
\begin{align}\label{G}
G^2_k&=\big|{\bf a}^{\mH}(\theta_0)({\bf w}_{k-1,\star}+{\bf T}^{-1}_{k-1}{\bf A}_k{\bf h}_{k})\big|^2\nonumber\\
&=-{\bf h}^{\mH}_{k}{\widetilde{\bf C}}{\bf h}_{k}+
2{\Re}\left({\widetilde{\bf c}}^{\mH}{\bf h}_{k}\right)+|{\bf a}^{\mH}(\theta_0){\bf w}_{k-1,\star}|^2
\end{align}
where $ {\bf w}_k={\bf T}^{-1}_k{\bf a}(\theta_0) $ is used, $ {\widetilde{\bf C}} $
and $ {\widetilde{\bf c}} $ are defined as
\begin{subequations}\label{qu187}
	\begin{align}
	{\widetilde{\bf C}}&\triangleq-({\bf T}^{-1}_{k-1}{\bf A}_k)^{\mH}{\bf a}(\theta_0){\bf a}^{\mH}(\theta_0)
	{\bf T}^{-1}_{k-1}{\bf A}_k
	\in{\mathbb C}^{{M_k}\times{M_k}}\\
	{\widetilde{\bf c}}&\triangleq({\bf T}^{-1}_{k-1}{\bf A}_k)^{\mH}{\bf a}(\theta_0){\bf a}^{\mH}(\theta_0){\bf w}_{k-1,\star}
	\in{\mathbb C}^{M_k}.
	\end{align}
\end{subequations}
Since
$ |{\bf a}^{\mH}(\theta_0){\bf w}_{k-1,\star}|^2 $ is a constant,
the maximization of $ G_k $ is thus equivalent to
the minimization of $ {\bf h}^{\mH}_{k}{\widetilde{\bf C}}{\bf h}_{k}-
2{\Re}\left({\widetilde{\bf c}}^{\mH}{\bf h}_{k}\right) $.

On the other hand, recalling the expression of $ L(\theta,\theta_0) $, we can rewrite
the constraint \eqref{cons1} as
\begin{align}\label{cons3}
{\bf w}^{\mH}_k{\bf S}_{k,m}{\bf w}_k=0,~m=1,\cdots,M_k
\end{align}
where $ {\bf S}_{k,m}={\bf a}(\theta_{k,m}){\bf a}^{\mH}(\theta_{k,m})-
{\rho}_{k,m}{\bf a}(\theta_{0}){\bf a}^{\mH}(\theta_{0}) $.
Substituting the constraint \eqref{cons} into \eqref{cons3}, we have
\begin{align}\label{cons7}
{\bf h}^{\mH}_{k}{\widetilde{\bf D}_m}{\bf h}_{k}-
2{\Re}({\widetilde{\bf d}}^{\mH}_m{\bf h}_{k})={\alpha}_m,~m=1,\cdots,M_k
\end{align}
where 
\begin{subequations}\label{qu188}
	\begin{align}
	{\widetilde{\bf D}_m}&=({\bf T}^{-1}_{k-1}{\bf A}_k)^{\mH}{\bf S}_{k,m}{\bf T}^{-1}_{k-1}{\bf A}_k
	\in{\mathbb C}^{{M_k}\times{M_k}}\\
	{\widetilde{\bf d}_m}&=-({\bf T}^{-1}_{k-1}{\bf A}_k)^{\mH}{\bf S}_{k,m}{\bf w}_{k-1,\star}
	\in{\mathbb C}^{M_k}\\
	{\alpha}_m&=-{\bf w}^{\mH}_{k-1,\star}{\bf S}_{k,m}{\bf w}_{k-1,\star}
	\in{\mathbb R}.
	\end{align}
\end{subequations}
Thus, problem \eqref{qu111} can be reformulated as
\begin{subequations}\label{qu163}
\begin{align}	
\min_{{\bf h}_{k}}&~~~{\bf h}^{\mH}_{k}{\widetilde{\bf C}}{\bf h}_{k}-
2{\Re}\left({\widetilde{\bf c}}^{\mH}
{\bf h}_{k}\right)\\
{\rm subject~to}&~~~{\bf h}^{\mH}_{k}{\widetilde{\bf D}_m}{\bf h}_{k}-
2{\Re}({\widetilde{\bf d}}^{\mH}_m{\bf h}_{k})={\alpha}_m\\
&~~~m=1,\cdots,M_k.\nonumber
\end{align}
\end{subequations}
In the sequel, we adopt the newly developed
C-ADMM approach \cite{conADMM} to solve problem \eqref{qu163}.

\subsubsection{C-ADMM Solver}
We first convert \eqref{qu163} into its real domain as
\begin{subequations}\label{qu169}	
	\begin{align}
	\min_{{\bf z}}&~~~{\bf z}^{\mT}{\bf C}{\bf z}-2{\bf c}^{\mT}{\bf z}\\
\label{qcqp1}{\rm subject~to}&~~~{\bf z}^{\mT}{\bf D}_m{\bf z}-
	2{\bf d}^{\mT}_m{\bf z}={\alpha}_m\\
	&~~~m=1,\cdots,M_k\nonumber
	\end{align}
\end{subequations}
where
\begin{subequations}\label{qu189}
	\begin{align}
\label{zz}{\bf z}&=\begin{bmatrix}{\Re}({\bf h}^{\mT}_{k}) & {\Im}({\bf h}^{\mT}_{k})\end{bmatrix}^{\mT}
	\in{\mathbb R}^{2{M_k}}\\
	{\bf c}&=\begin{bmatrix}{\Re}({\widetilde{\bf c}}^{\mT}) & {\Im}({\widetilde{\bf c}}^{\mT})\end{bmatrix}^{\mT}
	\in{\mathbb R}^{2{M_k}}\\
	{\bf d}_m&=\begin{bmatrix}{\Re}({\widetilde{\bf d}_m}^{\mT}) & {\Im}({\widetilde{\bf d}_m}^{\mT})\end{bmatrix}^{\mT}
	\in{\mathbb R}^{2{M_k}}\\
	{\bf C}&=\begin{bmatrix}{\Re}({\widetilde{\bf C}}) & -{\Im}({\widetilde{\bf C}})
	\\{\Im}({\widetilde{\bf C}}) & {\Re}({\widetilde{\bf C}})\end{bmatrix}
	\in{\mathbb R}^{2{M_k}\times 2{M_k}}\\
	{\bf D}_m&=\begin{bmatrix}{\Re}({\widetilde{\bf D}_m}) & -{\Im}({\widetilde{\bf D}_m})
	\\{\Im}({\widetilde{\bf D}_m}) & {\Re}({\widetilde{\bf D}_m})\end{bmatrix}
	\in{\mathbb R}^{2{M_k}\times 2{M_k}}.
	\end{align}
\end{subequations}

To tackle \eqref{qu169}, we introduce the auxiliary variable vectors $ {\bf p}_m $, $ m=1,\cdots,M_k $, and then formulate \eqref{qu169} as
\begin{subequations}\label{qu173}	
	\begin{align}
\label{li1near}\min_{{\bf z},\{{\bf p}_m\}^{M_k}_{m=1}}&~~~{\bf z}^{\mT}{\bf C}{\bf z}-2{\bf c}^{\mT}{\bf z}\\
	\label{linear}{\rm subject~to}&~~~{\bf p}_m={\bf z}\\
	\label{initialcons}&~~~{\bf p}^{\mT}_m{\bf D}_m{\bf p}_m-
	2{\bf d}^{\mT}_m{\bf p}_m={\alpha}_m\\
	&~~~m=1,\cdots,M_k.\nonumber
	\end{align}
\end{subequations}
Note that 
the non-convex constraint in problem \eqref{qu173} is only imposed on $ {\bf p}_m $
and not related to $ {\bf z} $.
Moreover, for any given $ m=1,\cdots,M_k $,
the nonconvex-constraint, i.e.,
\eqref{initialcons}, is a QCQP with only one constraint (QCQP-1), which can be easily solved as pointed out in \cite{conADMM}.
Thus, the newly formulated problem \eqref{qu173}
simplifies the original problem \eqref{qu169} to solve.

To see the details, we first devise the augmented Lagrangian
by ignoring the constraint \eqref{initialcons}:
\begin{align}\label{lag}
\mathcal{L}_{\eta}({\bf z},{\bf p},{\bm \lambda})=&
{\bf z}^{\mT}{\bf C}{\bf z}-2{\bf c}^{\mT}{\bf z}+\nonumber\\
&\sum_{i=1}^{M_k}{\bm \lambda}^{\mT}_m({\bf z}-{\bf p}_m)
+\sum_{i=1}^{M_k}\frac{\eta}{2}\|{\bf z}-{\bf p}_m\|^2_2
\end{align}
where $ \eta>0 $ is the penalty parameter, $ {\bm \lambda}_m\in{\mathbb R}^{2{M_k}} $
are Lagrange multiplier vectors.
Note that the augmented Lagrangian \eqref{lag}
acts as the (unaugmented) Lagrangian associated with the following
problem:
\begin{subequations}\label{qu1173}	
	\begin{align}
	\label{lnear1}\min_{{\bf z},\{{\bf p}_m\}^{M_k}_{m=1}}&~~~{\bf z}^{\mT}{\bf C}{\bf z}-2{\bf c}^{\mT}{\bf z}
	+\sum_{i=1}^{M_k}\frac{\eta}{2}\|{\bf z}-{\bf p}_m\|^2_2\\
	\label{linear1}{\rm subject~to}&~~~{\bf p}_m={\bf z},~m=1,\cdots,M_k
	\end{align}
\end{subequations}
which is equivalent to problem 
\eqref{li1near}-\eqref{linear}, since
for any feasible $ {\bf z} $ and $ {\bf p}_m $, $ m=1,\cdots,M_k $, the
added term, i.e., the last term in \eqref{lnear1}, to the objective function is zero.
As mentioned in \cite{admm1},
the augmented Lagrangian brings
robustness to the dual ascent method adopted later.

Since the constraints \eqref{initialcons} are imposed on $ {\bf p}_m $ and not related
to $ {\bf z} $, they only play
roles in finding $ {\bf p}_m $, $ m=1,\cdots,M_k $.
For this reason, we don't include
\eqref{initialcons} in the above augmented Lagrangian intentionally.
Instead, we take the constraints in \eqref{initialcons}
into consideration when minimizing 
$ \mathcal{L}_{\eta}({\bf z},{\bf p},{\bm \lambda}) $ as shown next.

The alternating direction method of multipliers (ADMM) \cite{admm1}, which 
is an operator splitting algorithm originally devised 
to solve convex optimization problems, has been 
explored as a heuristic method to solve non-convex problems \cite{conADMM}.
Following the decomposition-coordination procedure of 
ADMM in \cite{admm1}, we can determine $ \left\{{\bf z},{\bf p}_m,{\bm \lambda}\right\} $
via the alternative and iterative steps below.

\noindent
$ \bf Step~1 $: Update $ {\bf z} $
\begin{align}\label{key05}
{\bf z}(t+1)&={\rm arg}~\min_{\bf z}~\mathcal{L}_{\eta}({\bf z},{\bf p}(t),{\bm \lambda}(t))\nonumber\\
&={\rm arg}~\min_{\bf z}~{\bf z}^{\mT}({\bf C}+\frac{\eta{M}_k}{2}{\bf I}){\bf z}-2{\bf g}^{\mT}(t+1){\bf z}\nonumber\\
&=({\bf C}+\frac{\eta{M}_k}{2}{\bf I})^{-1}{\bf g}(t+1)
\end{align}
where $ {\bf g}(t+1)=
{\bf c}-(1/2)\sum\limits_{m=1}^{M_k}({\bm \lambda}_m(t)-{\eta}{\bf p}_m(t)) $.

\noindent
$ \bf Step~2 $: Update $ {\bf p} $

For $ m=1,\cdots,M_k $, we update the vector $ {\bf p}_m $ as
\begin{subequations}\label{key06}
	\begin{align}
\!\!{\bf p}_m(t+1)={\rm arg}~\min_{{\bf p}_m}&~\mathcal{L}_{\eta}({\bf z}(t+1),{\bf p},{\bm \lambda}(t))\nonumber\\
	={\rm arg}~\min_{{\bf p}_m}&~{\eta}{\bf p}^{\mT}_m{\bf p}_m-2({\eta}{\bf z}(t+1)+{\bm \lambda}_m(t))^{\mT}{\bf p}_m
	\nonumber\\
	={\rm arg}~\min_{{\bf p}_m}&~\|{\bf p}_m-{\bm \zeta}_m(t+1)\|^2_2\\
	{\rm subject~to}&~{\bf p}^{\mT}_m{\bf D}_m{\bf p}_m-
	2{\bf d}^{\mT}_m{\bf p}_m={\alpha}_m
	\end{align}	
\end{subequations}
where $ {\bm \zeta}_m(t+1)={\bf z}(t+1)+(1/{\eta}){\bm \lambda}_m(t) $.
Since the above problem is QCQP-1 which is equivalent to solving
a polynomial as mentioned in \cite{conADMM},
the bisection or Newton¡¯s method can be adopted to find its 
(approximate) solution, see \cite{conADMM} and \cite{qcqp} for reference.

\noindent
$ \bf Step~3 $: Update $ {\bm \lambda} $

For $ m=1,\cdots,M_k $, we update the vector $ {\bm \lambda}_m $ as
\begin{align}\label{criter}
{\bm \lambda}_m(t+1)={\bm \lambda}_m(t)+{\eta}({\bf z}(t+1)-{\bf p}_m(t+1)).
\end{align}

The above steps 1 to 3 are repeated until a stopping criterion is
reached, e.g., a maximum iteration number is attained and/or
\begin{align}\label{condi}
\delta>\delta_{\rm MAX}\triangleq\max\limits_{1\leq m\leq M_k}\|{\bf z}(t+1)-{\bf p}_m(t+1)\|_2
\end{align}
where $ \delta>0 $ is a small tolerance parameter.

\begin{algorithm}[!t]
	\caption{{C-ADMM Approach to Problem \eqref{qu101}}}\label{coded}
	\begin{algorithmic}[1]
		\State give $ {\bf a}(\theta_0) $, $ {\bf T}_{k-1} $, $ {\bf w}_{k-1,\star}={\bf T}^{-1}_{k-1}{\bf a}(\theta_0) $,
		$ \theta_{k,m} $, $ {\rho_{k,m}} $, $ m=1,\cdots,M_k $, and $ {\bf A}_k $, obtain $ {\bf c} $, $ {\bf C} $, $ {\bf d}_m $, $ {\bf D}_m $
		from \eqref{qu189}, initialize $ {\bf p}_m $, $ m=1,\cdots,M_k $, by \eqref{key36},
		set $ \delta_{\rm MAX}>\delta>0 $ and $ \eta>0 $		
		\While{$ \delta_{\rm MAX}>\delta $}
		\State update $ {\bf z} $ by \eqref{key05}
		\State update $ {\bf p}_m $, $ m=1,\cdots,M_k $, by \eqref{key06}
		\State update $ {\bm \lambda}_m $, $ m=1,\cdots,M_k $, by \eqref{criter}
		\State calculate $ \delta_{\rm MAX} $ by \eqref{condi}		
		\EndWhile
		\State obtain $ {\bf z}_{\star}={\bf z} $
		\State obtain $ {\bf h}_{k,\star} $	by \eqref{zz}
	\end{algorithmic}
\end{algorithm}

\subsubsection{Initialization of C-ADMM}
Note that due to the non-convexity of problem \eqref{qu173},
	typical convergence results on ADMM do not apply
	and the ultimate $ {\bf z} $ is not guaranteed to be optimal.
Nevertheless, an appropriate initialization 
makes the above iterative algorithm \cite{conADMM} work well and even converge to a 
Karush-Kuhn-Tucker (KKT) point.
Following \cite{conADMM}, we initialize $ {\bf p}_m $ as
\begin{align}\label{key35}
{\bf p}_m=\begin{bmatrix}{\Re}({\widetilde{\bf p}_m}^{\mT}) & {\Im}({\widetilde{\bf p}_m}^{\mT})\end{bmatrix}^{\mT},~m=1,\cdots,M_k
\end{align}
where
\begin{align}\label{key36}
{\widetilde{\bf p}_m}=[\underbrace{0,\cdots,0}_{m-1},{\gamma}_{m,\star},0,\cdots,0]^{\mT}\in\mathbb{C}^{M_k}.
\end{align}
In \eqref{key36}, $ {\gamma}_{m,\star} $ is obtained by the
OPARC algorithm and satisfies
\begin{align}
\dfrac{\big|({\bf w}_{k-1,\star}+{\gamma}_{m,\star}{\bf T}^{-1}_{k-1}{\bf a}(\theta_{k,m}))^{\mH}
{\bf a}(\theta_{k,m})\big|^2}
{\big|({\bf w}_{k-1,\star}+{\gamma}_{m,\star}{\bf T}^{-1}_{k-1}{\bf a}(\theta_{k,m}))^{\mH}
{\bf a}(\theta_{0})\big|^2}={\rho}_{k,m}.
\end{align}
It can be verified that, the constraints \eqref{initialcons} can be satisfied if the initial settings $ {\bf p}_m $, $ m=1,\cdots,M_k $, take \eqref{key35}.
This makes it easier to find an approximate solution of problem \eqref{qu173}.

Once the solution $ {\bf z}_{\star} $ has been obtained, we can reconstruct $ {\bf h}_{k,\star} $ by \eqref{zz} and obtain $ {\bf w}_{k,\star} $ as
\begin{align}\label{wehtk}
{\bf w}_{k,\star}
={\bf w}_{k-1,\star}+{\bf T}^{-1}_{k-1}{\bf A}_k{\bf h}_{k,\star}.
\end{align}
The INRs of the newly assigned virtual interferences can by calculated via
\begin{align}\label{hh}
\!\!\!{\bf \Sigma}_{k,\star}={\rm Diag}\left(-{\bf h}_{k,\star}\oslash
\left({\bf A}^{\mH}_k{\bf T}^{-1}_{k-1}\left(
{\bf a}(\theta_0)+{\bf A}_k{\bf h}_{k,\star}
\right)\right)\right).
\end{align}
To make the above procedure clear, we summarize the C-ADMM approach to solve problem
\eqref{qu101} in Algorithm \ref{coded}.
Notice from \cite{conADMM}
	that the C-ADMM approach is memory-efficient
	and can be implemented in a parallelized or distributed manner.
	Thus, for a large array, the C-ADMM approach in Algorithm \ref{coded} may
	be a better choice to solve problem \eqref{qu101} compared to the iterative approach in Algorithm \ref{cod1ed}, although more iterations may be needed.

\vspace*{-1\baselineskip}
\subsection{Update of Covariance Matrix}
Similar to the OPARC algorithm,
the VCM $ {\bf T}_k $ also needs to be renewed so as to facilitate
the next execution of multi-point OPARC.
From the above discussions, $ {\bf T}_k $ is updated as
\begin{align}\label{T}
{\bf T}_k={\bf T}_{k-1}+{\bf A}_k{\bf \Sigma}_{k,\star}{\bf A}^{\mH}_k.
\end{align}
Accordingly, the weight vector is 
\begin{align}\label{wopar}
{\bf w}_{k,\star}={\bf T}^{-1}_k{\bf a}(\theta_0).
\end{align}
This completes the procedure of multi-point OPARC.
Finally, we describe the steps of multi-point OPARC in Algorithm \ref{coded2}.

Note that in our proposed multi-point OPARC algorithm, we carry out the parameter determination in a
subspace with dimension $ M_k $, not in the whole space of dimension $ N $.
The benefit is the reduced amount of calculation.
In addition, one can see that at most $ M_{\rm max}=N-1 $ points
can be precisely controlled, due to the limited degrees of freedom in
problem \eqref{qu101} or \eqref{qu111}. 

As a remark, the differences between the recent $ {\textrm M}{\textrm A}^2\textrm{RC} $ in \cite{ref201} and the proposed multi-point OPARC in this paper are similar to those between 
$ {\textrm A}^2\textrm{RC} $ and OPARC described in \cite{p1} in details.

\begin{algorithm}[!t]
	\caption{Multi-point OPARC Algorithm}\label{coded2}
	\begin{algorithmic}[1]
		\State give $ {\bf a}(\theta_0) $, $ {\bf T}_{k-1} $ and the
		weight vector $ {\bf w}_{k-1,\star}={\bf T}^{-1}_{k-1}{\bf a}(\theta_0) $,
		prescribe the angle $ {\theta_{k,m}} $
		and the corresponding desired level $ {\rho_{k,m}} $, $ m=1,\cdots,M_k $
		\State calculate $ {\bf \Sigma}_{k,\star} $ or $ {\bf h}_{k,\star} $ using Algorithm \ref{cod1ed} or
		Algorithm \ref{coded}
		\State obtain $ {\bf T}_k $ by \eqref{T} and calculate $ {\bf w}_{k,\star} $ by \eqref{weightkk} or \eqref{wehtk}	
	\end{algorithmic}
\end{algorithm}

\section{Applications of Multi-point OPARC}
In this section, we present three applications of multi-point OPARC to
array signal processing.

\subsection{Array Pattern Synthesis}
Given the beam axis $ \theta_0 $, the problem of array pattern synthesis is
to find an appropriate $ N\times1 $ weight vector that makes the response $ L(\theta,\theta_0) $ meet some specific requirements.
For simplicity, we denote the desired pattern as $ L_d(\theta) $.
Basically, the proposed algorithm herein shares a similar concept of
pattern synthesis using $ {\textrm A}^2\textrm{RC} $ in \cite{snrf41}. However,
it is able to significantly reduce the number of iterations and improve
the performance.

\subsubsection{General Case}
Generally, the array pattern synthesis can be started by
setting $ k=0 $ and the initial weight as $ {\bf w}_{0,\star}={\bf a}(\theta_0) $.
For $ k>0 $,
multiple directions are selected by 
comparing $ L_{k-1}(\theta,\theta_0) $:
\begin{align}
L_{k-1}(\theta,\theta_0)\triangleq{|{\bf w}^{\mH}_{k-1}{\bf a}(\theta)|^2}
\big/{|{\bf w}^{\mH}_{k-1}{\bf a}(\theta_0)|^2}
\end{align}
with the desired pattern $ L_d(\theta) $ as follows. 
These angles can be in either the
sidelobe region or the mainlobe region.
For sidelobe synthesis, we only choose the peak angles in the set
\begin{align}
{\Omega}_{k,S}&=\left\{\theta\big|
L_{k-1}(\theta,\theta_0)>
L_{k-1}(\theta-\varepsilon,\theta_0){~\rm and}\right.\nonumber\\
&~~~~~~~\left.L_{k-1}(\theta,\theta_0)>
L_{k-1}(\theta+\varepsilon,\theta_0),~\theta\in\Omega_S\right\}
\end{align}
where $ \varepsilon $ is a small positive quantity, $\Omega_S$ denotes the sidelobe sector of the desired pattern.
Different from the angle selection method in $ {\textrm A}^2\textrm{RC} $
where the chosen peak angles have larger response levels than their desired
values,
a selected peak angle in set $ {\Omega}_{k,S} $ may have
a less response level than its desired one.
For mainlobe synthesis, some discrete angles where the responses deviate considerably from the desired ones are
chosen, and we denote the set of selected angles in the mainlobe region as $ {\Omega}_{k,M} $.
Then, we take:
\begin{align}\label{edd}
{\Omega}_k={\Omega}_{k,S}\cup{\Omega}_{k,M}
\triangleq\{\theta_{k,1},\cdots,\theta_{k,M_k}\}
\end{align}
where $ M_k={\rm card}({\Omega}_k) $.
The multi-point OPARC algorithm can thus be applied to adjust the 
corresponding responses of angles $ \theta_{k,m} $ to their desired
values $ \rho_{k,m}=L_d(\theta_{k,m}) $, $ m=1,\cdots,M_k $, and 
the current response pattern $ L_k(\theta,\theta_0) $ can be
obtained by using the resulting weight of multi-point OPARC.
Then, set $ k=k+1 $ and repeat the above process
until the response is satisfactorily synthesized.
Note that the above iteration procedure is different from that
in Section II.C where $ k $ is fixed and an internal iteration 
within the $ k $th step is conducted.
To summarize, we describe the multi-point OPARC based array pattern synthesis algorithm
in Algorithm \ref{co1ded2}.
As mentioned earlier, $ {\Omega}_k $ is forced to satisfy
$ {\rm card}({\Omega}_k)<N $.
Otherwise, we can simply reduce $ {\rm card}({\Omega}_k) $ by modifying $ {\Omega}_k $
similar to what is done next.

\subsubsection{Particular Consideration for Large Arrays}
As aforementioned, the proposed multi-point OPARC algorithm
operates in an $ M_k $-dimensional subspace of the original
$ N $-dimensional space.
This provides us an effective strategy to pattern synthesis for
large arrays, where the traditional approaches may not work well
or require extensive computation due to the large dimension.

More specifically, for a large array and a pre-determined angle set $ {\Omega}_k $ (whose cardinality
normally approaches to $ N $) in \eqref{edd},
we construct a new angle set $ {\Theta}_k $ as
\begin{align}\label{theta}
{\Theta}_k=
\left\{{\bar\theta}_{k,1},
{\bar\theta}_{k,2},\cdots, {\bar\theta}_{k,C_k}\right\}
\end{align}
where $ C_k $ is a prescribed number that is much smaller than $ N $,
$ {\bar\theta}_{k,c} $, $ c=1,\cdots,C_k $, is the $ c $th element of the
vector:
\begin{align}\label{theta1}
{\rm Sort}({\Omega}_k)\in\mathbb{R}^{{\rm card}({\Omega}_k)}
\end{align}
where $ {\rm Sort}({\Omega}_k) $ re-arranges the
elements of $ {\Omega}_k $ in the following way:
the larger $ |L_{k-1}({\bar\theta},\theta_0)-L_d({\bar\theta})| $
for $ {\bar\theta}\in{\Omega}_k $ is, 
the smaller index of $ {\bar\theta} $ in $ {\rm Sort}({\Omega}_k) $ is, which makes
$ {\bar\theta} $ more likely to be chosen as an element in the angle set
$ {\Theta}_k $ in \eqref{theta}.
The reason for this is that we expect to reduce the overall difference
between the resulting pattern and the desired one.

Once the new angle set $ {\Theta}_k $ is obtained, the multi-point OPARC algorithm can
be applied to realize
$ L_k({\bar\theta},\theta_0)=L_d({\bar\theta}) $
for $ {\bar\theta}\in{\Theta}_k $. 
Then, set $ k=k+1 $ and repeat the above process until the response is satisfactorily synthesized, and the cardinality of set $ {\Theta}_k $, i.e., $ C_k $, 
can be flexibly varied with the iteration number $ k $.
Finally, the above-described large-array pattern synthesis
can be readily realized via Algorithm \ref{co1ded2},
by simply replacing $ {\Omega}_k $ in the 4th line of Algorithm \ref{co1ded2} with the
new angle set $ {\Theta}_k $ in \eqref{theta}.

Since the above proposed algorithm, in either the general case or
	the large-array scenario, iteratively adjusts
	the responses of sidelobe peaks, it
	is able to make all the sidelobe peaks align with
the desired values. Thus, all the sidelobe responses
can be well controlled to be lower than the given thresholds,
and a satisfactory sidelobe shape can be well maintained.
Nevertheless, array pattern synthesis works in a data-independent way,
the resulting weight or its corresponding beampattern is
lack of adaptivity in suppressing undesirable interference and
noise, which can be well rejected by the adaptive beamformer as discussed next.

\begin{algorithm}[!t]
	\caption{Multi-point OPARC based Array Pattern Synthesis Algorithm}\label{co1ded2}
	\begin{algorithmic}[1]
		\State give $ L_d(\theta) $, $ {\bf w}_{0,\star}={\bf a}(\theta_0) $,
		set $ k=1 $, $ {\bf T}_0={\bf I} $, $ \varepsilon>0 $
		\While{$ 1 $}
		\State determine $ {\Omega}_k $ from \eqref{edd}
		\State apply multi-point OPARC algorithm to realize
		$ ~~~~~L_k(\theta,\theta_0)=L_d(\theta) $ ($ \theta\in{\Omega}_k $),
		update $ {\bf w}_{k,\star} $ and $ {\bf T}_{k} $
		\If{$ L_k(\theta,\theta_0) $ meets the requirement}
		\State break
		\EndIf	
		\State set $ k=k+1 $	
		\EndWhile
		\State output $ {\bf w}_{k,\star} $	and $ L_k(\theta,\theta_0) $
	\end{algorithmic}
\end{algorithm}

\subsection{Multi-constraint Adaptive Beamforming}
The linearly constrained minimum variance (LCMV) beamformer is commonly used
to enhance the robustness of array systems \cite{lcmv,lcmv2,lcmv3}.
In LCMV beamformer, several linear constraints are imposed when minimizing the output
variance, i.e.,\begin{subequations}\label{lcmv}	
	\begin{align}
	\min_{\bf w}&~~~{\bf w}^{\mH}{\bf R}_{n+i}{\bf w}\\
	\label{lcmv2}{\rm subject~to}&~~~{\bf C}^{\mH}{\bf w}={\bf g}
	\end{align}
\end{subequations}
where $ {\bf C} $ is the constraint matrix that consists of $ D $ spatial
steering vectors corresponding to the $ D $ constrained directions 
$ \theta_d $, $ d=0,\cdots,D-1 $, i.e.,
$ {\bf C}=[{\bf a}(\theta_0), {\bf a}(\theta_1), \cdots, {\bf a}(\theta_{D-1})] $,
$ {\bf g} $ is a prescribed $ D $-dimensional vector usually satisfying
$ ({\bf g})_1=1 $.
The solution of problem \eqref{lcmv} is given by
\begin{align}\label{lcmvw}
{\bf w}_{\rm LCMV}={\bf R}^{-1}_{n+i}{\bf C}
({\bf C}^{\mH}{\bf R}^{-1}_{n+i}{\bf C})^{-1}{\bf g}.
\end{align}

From \eqref{lcmv2}, we can clearly see that both the amplitude and the phase of
the array output, i.e., $ {\bf w}^{\mH}{\bf a}(\theta) $,
have been strictly constrained at $ \theta_d $, $ d=0,\cdots,D-1 $.
As a matter of fact, a less restrictive quadratically constrained minimum variance (QCMV) beamformer
should be formulated by removing the unnecessary phase constraints, i.e.,
\begin{subequations}\label{QCMV}	
	\begin{align}
	\label{QCMV1}\min_{\bf w}&~{\bf w}^{\mH}{\bf R}_{n+i}{\bf w}\\
	\label{QCMV2}{\rm subject~to}&~|({\bf C}^{\mH}{\bf w})_d|^2=|({\bf g})_d|^2,~d=1,\cdots,D.
	\end{align}
\end{subequations}
Note that in this subsection the variable $ d $ 
is an index and does not mean ``desired'' as used previously.
Comparing to the QCMV in \eqref{QCMV}, we can see that the LCMV beamformer
in \eqref{lcmv} strictly limits the optimization of the
weight vector to a smaller space, although it has a closed-from solution.
It, thus, may cause the output 
SINR of LCMV beamformer to suffer from a loss, and the resulting pattern may be distorted.

We adopt the multi-point OPARC algorithm to solve 
the QCMV problem \eqref{QCMV}, in the hope that the resulting output SINR can be improved
(comparing to LCMV).
If $ D=1 $, i.e., one constraint $ |{\bf a}^{\mH}(\theta_0){\bf w}|^2=1 $
is imposed in \eqref{QCMV2}, the optimal solution of \eqref{QCMV} is given by
\begin{align}\label{key}
{\bf w}=\dfrac{{\bf R}^{-1}_{n+i}{\bf a}(\theta_0)}
{{\bf a}^{\mH}(\theta_0){\bf R}^{-1}_{n+i}{\bf a}(\theta_0)}.
\end{align}
If $D>1$, based on the first constraint that 
$ |{\bf a}^{\mH}(\theta_0){\bf w}|^2=1 $, 
we have $ L(\theta_{d-1},\theta_0)={|{{\bf w}^{\mH}}{\bf a}(\theta_{d-1})|^2} $
in \eqref{QCMV2}.
Then, the additional $ (D-1) $ constraints can be taken into account by imposing the following constraints:
\begin{align}\label{re}
L(\theta_{d-1},\theta_0)=|({\bf g})_d|^2,~d=2,\cdots,D.
\end{align}
Then, the problem becomes how to realize the above described
multi-point response control, starting from the optimal weight
vector in \eqref{key}.
To apply the multi-point OPARC algorithm, we rewrite $ {\bf w} $ in \eqref{key} as
\begin{align}
{\bf w}=\dfrac{1}
{{\sigma}^2_{n}{\bf a}^{\mH}(\theta_0){\bf R}^{-1}_{n+i}{\bf a}(\theta_0)}
{\bf T}^{-1}_{n+i}{\bf a}(\theta_0)\triangleq c{\bf w}_{0}
\end{align}
where $ c $ is a constant satisfying $ c=({\sigma}^2_{n}{\bf a}^{\mH}(\theta_0){\bf R}^{-1}_{n+i}{\bf a}(\theta_0))^{-1} $, 
$ {\bf T}_{n+i} $ and $ {\bf w}_{0}={\bf T}^{-1}_{n+i}{\bf a}(\theta_0) $ 
act as the initial VCM in \eqref{part201} and the initial
weight vector in multi-point OPARC, respectively.
Then,
a multi-point OPARC procedure can be applied to fulfill the
response requirement described in \eqref{re},
and the ultimate weight vector of QCMV (denoted as $ {\bf w}_{\rm QC} $) can be
obtained accordingly.

\begin{figure*}[t]
	\normalsize
	\setcounter{MYtempeqncnt}{\value{equation}}
	\setcounter{equation}{53}
	\vspace*{-2\baselineskip}	
	\begin{align}\label{key2}
	{{\bf w}^{\mH}_{\rm QC}}{\bf T}_{\rm QC}{\bf w}_{\rm QC}=
	{{{\bf w}^{\mH}_{\rm QC}}\left({\bf T}_{n+i}+
		{\sum\limits_{d=2}^{D}}{\beta_{d-1}}{\bf a}(\theta_{d-1}){\bf a}^{\mH}(\theta_{d-1})\right)
		{\bf w}_{\rm QC}}
	&=
	{{\bf w}^{\mH}_{\rm QC}}{\bf T}_{n+i}{\bf w}_{\rm QC}
	+\underbrace{|{{\bf w}^{\mH}_{\rm QC}}{\bf a}(\theta_0)|^2
		{
			\left({\sum\limits_{d=2}^{D}}
			{{\beta_{d-1}}|({\bf g})_d|^2}\right)}}_{=0}
	\nonumber\\
	&={{\bf w}^{\mH}_{\rm QC}}{\bf T}_{n+i}{\bf w}_{\rm QC}=
	\dfrac{
	{{\bf w}^{\mH}_{\rm QC}}{\bf R}_{n+i}{\bf w}_{\rm QC}}
	{\sigma^2_n}
	\end{align}
	\setcounter{equation}{\value{MYtempeqncnt}}
	\hrulefill
	\vspace*{-1\baselineskip}
\end{figure*}

Note that in practical applications, $ {\bf R}_{n+i} $ 
can be estimated from data $ {\bf x}(t) $:
\begin{align}\label{estR}
{\hat {\bf R}}_{n+i}=\dfrac{1}{T}\sum_{t=1}^{T}{\bf x}(t){\bf x}^{\mH}(t)
\end{align}
where $ T $ is the number of snapshots. 
In addition, $ {\sigma}^2_n $ can be estimated by \cite{sigma}
\begin{align}\label{estSigma}
{\hat{\sigma}}^2_n=\dfrac{1}{N-J_r}\sum_{n=J+1}^{N}\lambda_n
\end{align}
where $ J_r $ is the number of interferences, $ \lambda_1\geq\lambda_2\geq\cdots\geq\lambda_N $
are eigenvalues of $ {\hat {\bf R}}_{n+i} $.
Replacing $ {\bf R}_{n+i} $ and $ {\sigma}^2_n $ with
$ {\hat {\bf R}}_{n+i} $ and $ {\hat{\sigma}}^2_n $, respectively,
we have summarized the proposed algorithm in Algorithm \ref{co3ded2}.

To have a better understanding, we denote the corresponding VCM of $ {\bf w}_{\rm QC} $ as
$ {\bf T}_{\rm QC} $. Recalling the property \eqref{wopar} of multi-point OPARC, 
$ {\bf w}_{\rm QC} $ and $ {\bf T}_{\rm QC} $ satisfy
\begin{align}\label{keyw}
{\bf w}_{\rm QC}={\bf T}^{-1}_{\rm QC}{\bf a}(\theta_0).
\end{align}
We can see that the obtained weight $ {\bf w}_{\rm QC} $ minimizes the total variance $ {\bf w}^{\mH}{\bf T}_{\rm QC}{\bf w} $
with the constraints \eqref{QCMV2}, rather than minimizing $ {\bf w}^{\mH}{\bf T}_{n+i}{\bf w} $ or its equivalent term
$ {\bf w}^{\mH}{\bf R}_{n+i}{\bf w} $ (for a fixed $ \sigma^2_n $) in \eqref{QCMV1}.
Nevertheless, we know from Proposition 7 of the companion paper \cite{p1} that
the obtained weight of OPARC also minimizes
the variance at the previous step. 
Thus,
$ {\bf w}_{\rm QC} $ is the optimal solution of problem \eqref{QCMV}
for the special case when $ D=2 $, i.e.,
only one extra constraint is 
imposed besides the constraint $ |{\bf a}^{\mH}(\theta_0){\bf w}|^2=1 $.
In addition, the obtained $ {\bf w}_{\rm QC} $ offers the optimal solution of problem 
\eqref{QCMV} if we impose null constraint at $ \theta_{d-1} $, $ d=2,\cdots,D $,
based on the following argument.
In this case, we set $ |({\bf g})_d|^2=0 $, $ d=2,\cdots,D $, and thus obtain
\eqref{key2} on the top of this page,
where we have used the fact that
\addtocounter{equation}{1}
\begin{align}
\dfrac{|{{\bf w}^{\mH}_{\rm QC}}{\bf a}(\theta_{d-1})|^2}{|{{\bf w}^{\mH}_{\rm QC}}{\bf a}(\theta_0)|^2}=
|({\bf g})_d|^2=0,~d=2,\cdots,D
\end{align}
and
\begin{align}\label{keyT}
{\bf T}_{\rm QC}={\bf T}_{n+i}+
{\sum\limits_{d=2}^{D}}{\beta_{d-1}}{\bf a}(\theta_{d-1}){\bf a}^{\mH}(\theta_{d-1})
\end{align}
with $ \beta_{d-1} $ denoting the
INR of the assigned virtual interference at $ \theta_{d-1} $.
From \eqref{key2} we know that $ {{\bf w}_{\rm QC}} $ also
minimizes $ {\bf w}^{\mH}{\bf R}_{n+i}{\bf w} $.
The optimality (in the sense of output SINR) of the proposed algorithm is 
guaranteed in the above two scenarios.
Otherwise, the proposed algorithm performs better than LCMV algorithm
in most cases
as we shall see from the simulations later.

\begin{algorithm}[!t]
	\caption{Multi-point OPARC based Multi-constraint Adaptive Beamforming Algorithm}\label{co3ded2}
	\begin{algorithmic}[1]
		\State give interference number $ J_r $, constraint matrix $ \bf C $ and vector 
		$ {\bf g} $, estimate $ {\hat {\bf R}}_{n+i} $ and $ {\hat{\sigma}}^2_n $ by \eqref{estR} and \eqref{estSigma}, respectively,
		calculate $ {\bf T}_{n+i}={\hat {\bf R}}_{n+i}/{\hat{\sigma}}^2_n $ and 
		$ {\bf w}_{0}={\bf T}^{-1}_{n+i}{\bf a}(\theta_0) $
		\State apply multi-point OPARC algorithm to realize
		$ L(\theta_{d-1},\theta_0)=|({\bf g})_d|^2 $, $ d=2,\cdots,D $, by setting
		$ {\bf T}_{n+i} $ and $ {\bf w}_{0} $ as the initial VCM and the initial weight vector, respectively, to obtain $ {\bf w}_{\rm QC} $
	\end{algorithmic}
\end{algorithm}

Moreover, \eqref{keyw} and \eqref{keyT} indicate that the resulting
weight vector $ {\bf w}_{\rm QC} $ is obtained by making a normalized covariance 
matrix loading (NCL), which can be regarded as a generalization of the diagonal loading (DL) in \cite{ld1,ld2,ld3}, 
on the initial $ {\bf T}_{n+i} $. The loading quantity
is precisely determined by multi-point OPARC algorithm as
\begin{align}\label{ldd}
{\bf \Delta}={\sum_{d=2}^{D}}{\beta_{d-1}}{\bf a}(\theta_{d-1}){\bf a}^{\mH}(\theta_{d-1}).
\end{align}
Recalling Eqn. (38) of \cite{p1}, one learns in OPARC that the INR of
	a newly assigned virtual interference depends on the 
	previous normalized covariance matrix and also contributes to the
	current one.
Then, revisiting Algorithm \ref{cod1ed}, where OPARC is iteratively applied, and
Eqn. \eqref{INR1}, one can see that the resulting $ \beta_{d-1} $, $ d=2,\cdots,D $,
depend on the initial $ {\bf T}_{n+i} $.
Thus, the loading quantity $ {\bf \Delta} $
in \eqref{ldd} is related to the given constraints in \eqref{QCMV2} and also the 
real data.

Note that the above-described multi-constraint adaptive beamforming algorithm
improves the robustness of array systems while blocking
the unexpected interference and noise.
However, different from the method in the 
	preceding subsection where the sidelobe peaks can be controlled iteratively, 
	the algorithm in this subsection only has constraints on 
	the response levels of several pre-assigned angles
	$ \theta_0,\theta_1,\cdots,\theta_{D-1} $. 
	It cannot control/guarantee an overall sidelobe pattern.

\vspace*{-0.6\baselineskip}
\subsection{Quiescent Pattern Control}
In adaptive beamforming, weight vector is designed in a data-dependent manner.
However, the traditional adaptive beamforming methods
usually yield a beampattern with high sidelobes.
To obtain low sidelobes in adaptive arrays,
the concept of quiescent pattern
control is introduced in \cite{con6},
by combining the adaptive beamforming and deterministic
pattern synthesis techniques.
In brief, when an adaptive array operates in the presence of white noise only,
the resultant adaptive beamformer is named as the quiescent
weight vector, and the corresponding array response is termed as the
quiescent pattern.
Following the concept of quiescent pattern control in \cite{con6,con200,con1}, 
it is required to find a mechanism 
to design a beamformer having the ability
to reject an interference (if it exists) and noise, and meanwhile, 
maintaining the desirable
shape of the quiescent pattern when only white noise presents.

Note that the quiescent weight vector of LCMV beamformer in \eqref{lcmvw} is
${\bf w}_{q}={\bf C}({\bf C}^{\mH}{\bf C})^{-1}{\bf g}$ that 
can be readily obtained by setting $ {\bf R}_{n+i}=\sigma^2_n{\bf I} $.
Unfortunately, for a given desired quiescent pattern, 
which usually has specific constraints on 
the upper level of sidelobes,
it is not easy
to have a satisfactory quiescent pattern via LCMV by
specifying $ {\bf C} $ and $ {\bf g} $, since
LCMV only imposes constraints on a fixed set of 
	pre-assigned  finite angles as mentioned at the end of Section III.B.
This is similarly true for the multi-point OPARC algorithm 
	presented in the preceding Section III.B.
Moreover, if we employ the iterative approach adopted in deterministic
	pattern synthesis in Section III.A to modify the shape of the obtained beampattern,
nulls may not be always formed at the directions of unknown real interferences,
and the adaptivity in suppressing undesirable components is thus
not well guaranteed.


In this subsection, a systematic approach to quiescent pattern control is 
proposed. A two-stage procedure is developed,
by taking advantage of the deterministic
pattern synthesis approach in Section III.A and also the concept of NCL
mentioned in Section III.B. 
More specifically, given a desired quiescent pattern, denoted as $ L_d(\theta) $,
	the multi-point OPARC based pattern synthesis algorithm in Section III.A, see, 
Algorithm \ref{co1ded2}, is adopted in the first stage to design
a desirable quiescent pattern off-line.
Denote by $ {\bf w}_q $, $ {\bf T}_q $ and $ L_q(\theta,\theta_0) $ the obtained 
(quiescent) weight vector, 
the associated VCM and the resulting response pattern, respectively.
It satisfies
\begin{align}\label{wq}
{\bf w}_q={\bf T}^{-1}_q{\bf a}(\theta_0).
\end{align}
As mentioned earlier, 
	the resulting $ L_q(\theta,\theta_0) $ performs well in maintaining
	the shape of $ L_d(\theta) $, however, the above weight $ {\bf w}_q $ has no
ability to reject the potential interferences and noise.
A strategy of finding
weight vector is thus required in quiescent pattern control to,
not only maintain the shape of $ L_d(\theta) $ if only white noise exists,
but also suppress a possible 
real interference and noise.
From the adaptive array theory, a data-dependent 
loading quantity $ {\bf \Delta} $ needs to be added to
the VCM $ {\bf T}_q $, such that the potential interferences and noise can be rejected. 
Moreover,
in the white noise only case,
$ {\bf \Delta} $ should be zero
such that the weight $ {\bf w}_q $ in \eqref{wq} can be retrieved.
To do so, we carry out the second stage, by taking a real data into consideration
and carrying out an NCL operator to the VCM $ {\bf T}_q $ via 
setting the associated loading quantity $ {\bf \Delta} $ as
\begin{align}
{\bf \Delta}=-{\bf I}+{\bf T}_{n+i}
\end{align}
where $ {\bf T}_{n+i}={\bf R}_{n+i}/{\sigma^2_n} $.
The ultimate (adaptive) weight vector is thus calculated as
\begin{align}\label{keya}
{\bf w}_{a}=({\bf T}_q-{\bf I}+{\bf T}_{n+i})^{-1}{\bf a}(\theta_0).
\end{align}
The corresponding response pattern of $ {\bf w}_{a} $ (denoted as $ L_a(\theta,\theta_0) $) 
can be obtained accordingly.

One can see that
there are two components being suppressed by $ {\bf w}_a $ in \eqref{keya}.
The first one is the component of the virtual interference which corresponds to
$ {\bf T}_q-{\bf I} $ and helps to maintain the shape of $ L_d(\theta) $. 
The second component is $ {\bf T}_{n+i} $, which
	contains the real interference and noise that
need to be rejected.
In the noise only scenario,
the loading quantity $ {\bf \Delta} $ offsets zero automatically
and the quiescent weight vector $ {\bf w}_q $ in \eqref{wq}
appears, provided that the real noise shares the same
structure as the virtual noise, i.e., $ {\bf R}_{n+i}={\sigma}^2_n{\bf I} $
or $ {\bf T}_{n+i}={\bf I} $.
Therefore, we can see that the weight vector $ {\bf w}_q $ 
in \eqref{wq} and its corresponding beampattern $ L_q(\theta,\theta_0) $
are exactly the quiescent weight vector and quiescent pattern, respectively.
Also, we should replace the unknown $ {\bf R}_{n+i} $ and $ \sigma^2_n $
with $ {\hat {\bf R}}_{n+i} $ in \eqref{estR} and $ {\hat{\sigma}}^2_n $ in \eqref{estSigma},
respectively, 
and set $ {\bf T}_{n+i}={\hat {\bf R}}_{n+i}/{\hat{\sigma}}^2_n $
in practical applications.

\begin{algorithm}[!t]
	\caption{Multi-point OPARC based Quiescent Pattern Control Algorithm}\label{co6ded2}
	\begin{algorithmic}[1]
		\State give $ L_d(\theta) $, synthesize a desirable quiescent pattern $ L_q(\theta,\theta_0) $ using Algorithm \ref{co1ded2}, obtain $ {\bf w}_q $ and $ {\bf T}_q $
		\State estimate $ {\hat {\bf R}}_{n+i} $ and $ {\hat{\sigma}}^2_n $ by \eqref{estR} and \eqref{estSigma}, respectively,	set $ {\bf T}_{n+i}={\hat {\bf R}}_{n+i}/{\hat{\sigma}}^2_n $
		\State obtain adaptive weight vector $ {\bf w}_a $ by Eqn. \eqref{keya}
		\State if extra constraints needed, modify $ {\bf w}_a $ by
		conducting the multi-point OPARC algorithm
		\State output the obtained weight $ {\bf w}_a $ and its corresponding response
		pattern $ L_a(\theta,\theta_0) $
	\end{algorithmic}
\end{algorithm}

It should be emphasized that we do not impose extra constraints
(e.g., fixed null constraints considered in \cite{con6})
on the resulting response pattern $ L_a(\theta,\theta_0) $, since such kind of constraints can be aforehand considered in the first stage of the above procedure.
In addition, we can also make the fixed constraints satisfied
by performing the multi-point OPARC algorithm 
starting from the obtained $ {\bf w}_a $ in \eqref{keya} and 
its corresponding normalized covariance matrix
$ {\bf T}={\bf T}_q-{\bf I}+{\bf T}_{n+i} $.
This is similar to the idea used in the preceding subsection.
To make it clear, we have summarized the multi-point OPARC based quiescent pattern control algorithm
in Algorithm \ref{co6ded2}.

\vspace*{-0.5\baselineskip}
\section{Numerical Results}
We next present some simulations to demonstrate
the proposed multi-point OPARC algorithm and its applications.
Unless otherwise specified,
we set $ \omega=6\pi\times10^{8}~{\rm rad/s} $ and
consider an 11-element nonuniform spaced linear array
with nonisotropic elements.
Both the element locations $ x_n $ and the element patterns $ g_n(\theta) $ are listed in 
Table I in Part I \cite{p1}, and the same array configuration has been adopted in Part I \cite{p1}.
The beam axis is steered to $ \theta_0=20^{\circ} $.
We set $ {\beta}_{\epsilon}=10^{-10} $
in conducting the iterative approach, and
take $ \delta=10^{-15} $ and $ \eta=900 $ for the C-ADMM approach.
In addition, $ {\bf f}_{n} $ is specified as
the all-zero vector for the ${\textrm {MA}}^2{\textrm{RC}}$ algorithm
in \cite{ref201} for comparison,
SNR is taken as $ 10{\rm dB} $ when it applies.

\vspace*{-0.5\baselineskip}
\subsection{Illustration of Multi-point OPARC}
In this subsection, we demonstrate the
multi-point OPARC algorithm.
Both the iterative approach and the C-ADMM approach are conducted, and then compared
with the ${\textrm M}{\textrm A}^2{\textrm{RC}}$ algorithm.
For convenience, we carry out two steps of the array response control algorithms 
with each step controlling two angles, i.e., $ M_1=M_2=2 $, and
denote the adjusted angles and the corresponding desired levels of the $ k $th ($ k=1,2 $) step as 
$ \theta_{k,m} $ and $ {\rho}_{k,m} $, $ m=1,\cdots, M_k $, respectively.
Following the evaluation strategy adopted in \cite{p1}, we
define
\begin{align}
D_m\triangleq|L_2(\theta_{1,m},\theta_0)-L_1(\theta_{1,m},\theta_0)|
\end{align}
to measure the response level differences between two consecutive response controls at 
$ \theta_{1,m} $, $ m=1,\cdots,M_1 $,
where $ L_k(\theta,\theta_0) $ represents the resultant response after finishing
the $ k $-th step of weight update, $ k=1, 2 $.
In addition, the deviation $ J $:
\begin{align}
J\buildrel \Delta \over =\sqrt{\dfrac{1}{I} \sum\limits^{I}_{i=1}
	\begin{vmatrix}L_2(\vartheta_i,\theta_0)-L_1(\vartheta_i,\theta_0)\end{vmatrix}^2}
\end{align}
is also considered,
where $ \vartheta_i $ stands for the $ i $th sampling point in the angle sector,
$ I $ denotes the number of sampling points.

\begin{figure}[!t]
	\centering
	\includegraphics[width=3.13in]{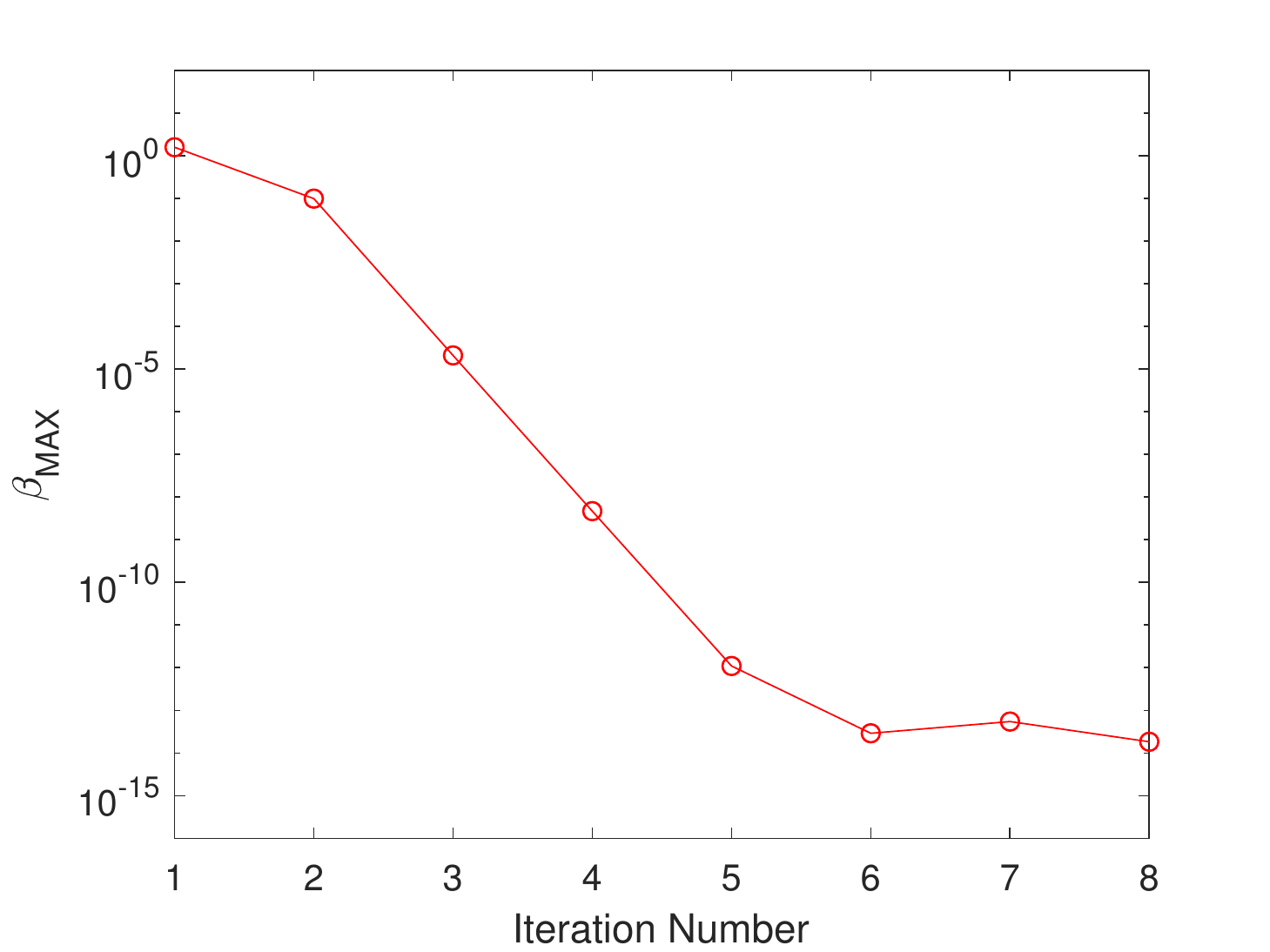}
	\caption{Curve of $ \beta_{\rm MAX} $ versus the iteration number.}
	\label{beta}
\end{figure}
\begin{figure}[!t]
	\centering
	\includegraphics[width=3.13in]{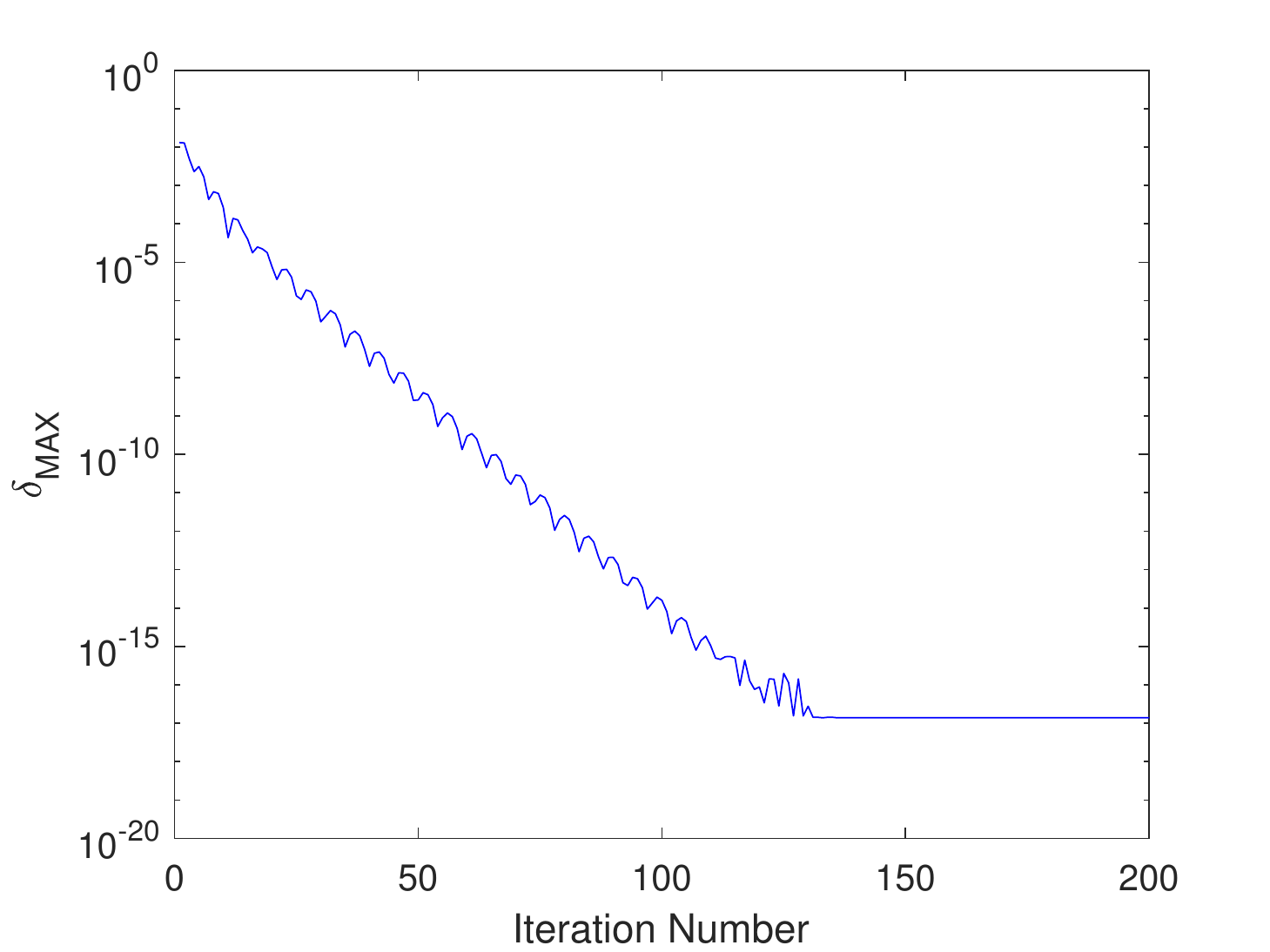}
	\caption{Curve of $ \delta_{\rm MAX} $ versus the iteration number.}
	\label{delta}
\end{figure}

\begin{figure}[!tpb]
	\centering\subfloat[The first step]
	{\includegraphics[width=3.13in]{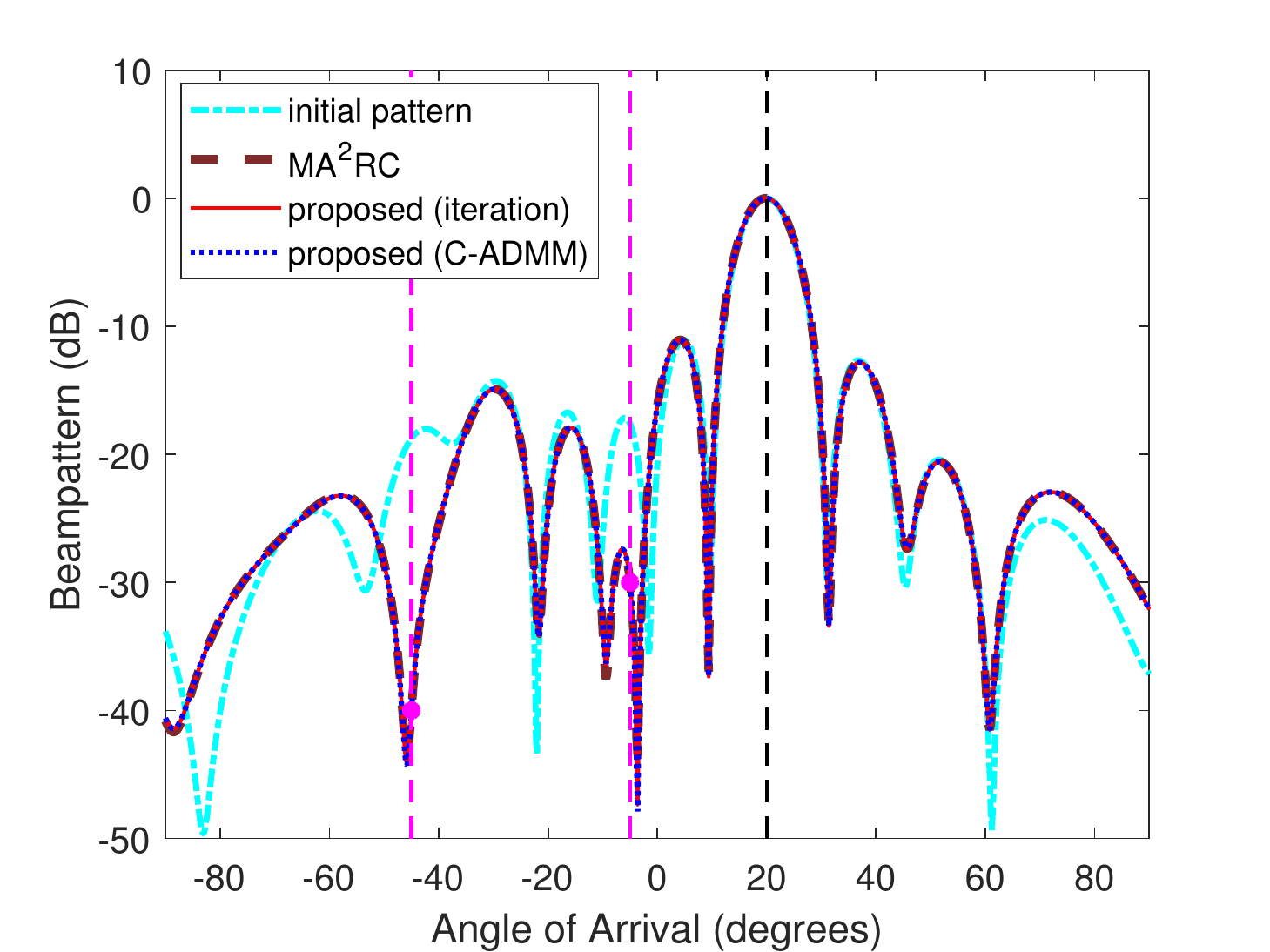}%
		\label{p3control}}\
	\centering\subfloat[The second step]
	{\includegraphics[width=3.13in]{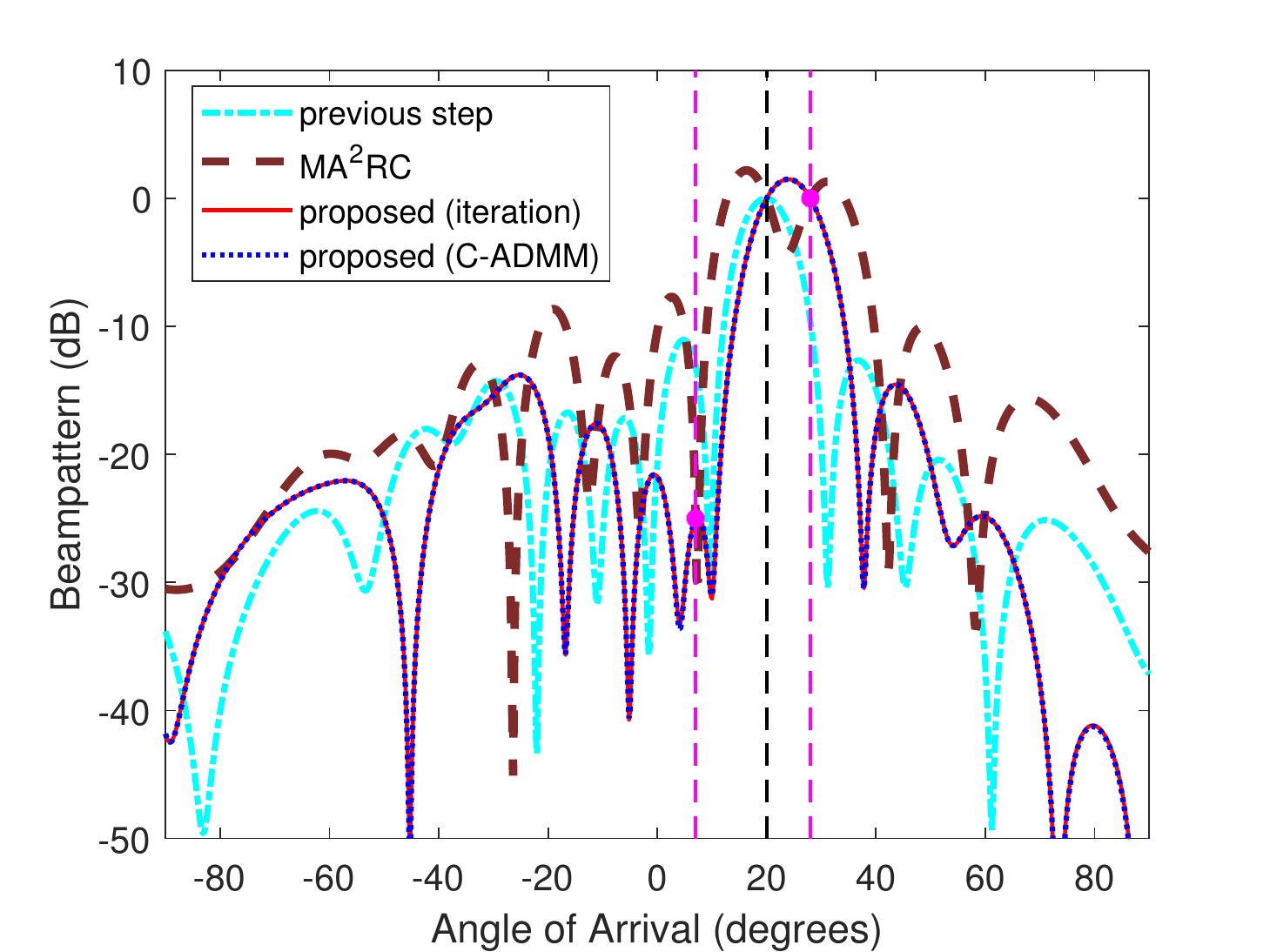}%
		\label{p3control2}}			
	\caption{Illustration of multi-point OPARC algorithm.}
	\label{dq1ddd}
\end{figure}

\begin{figure*}[!t]
	\centering
	\subfloat[Synthesized pattern at the 1st step]
	{\includegraphics[width=2.35in]{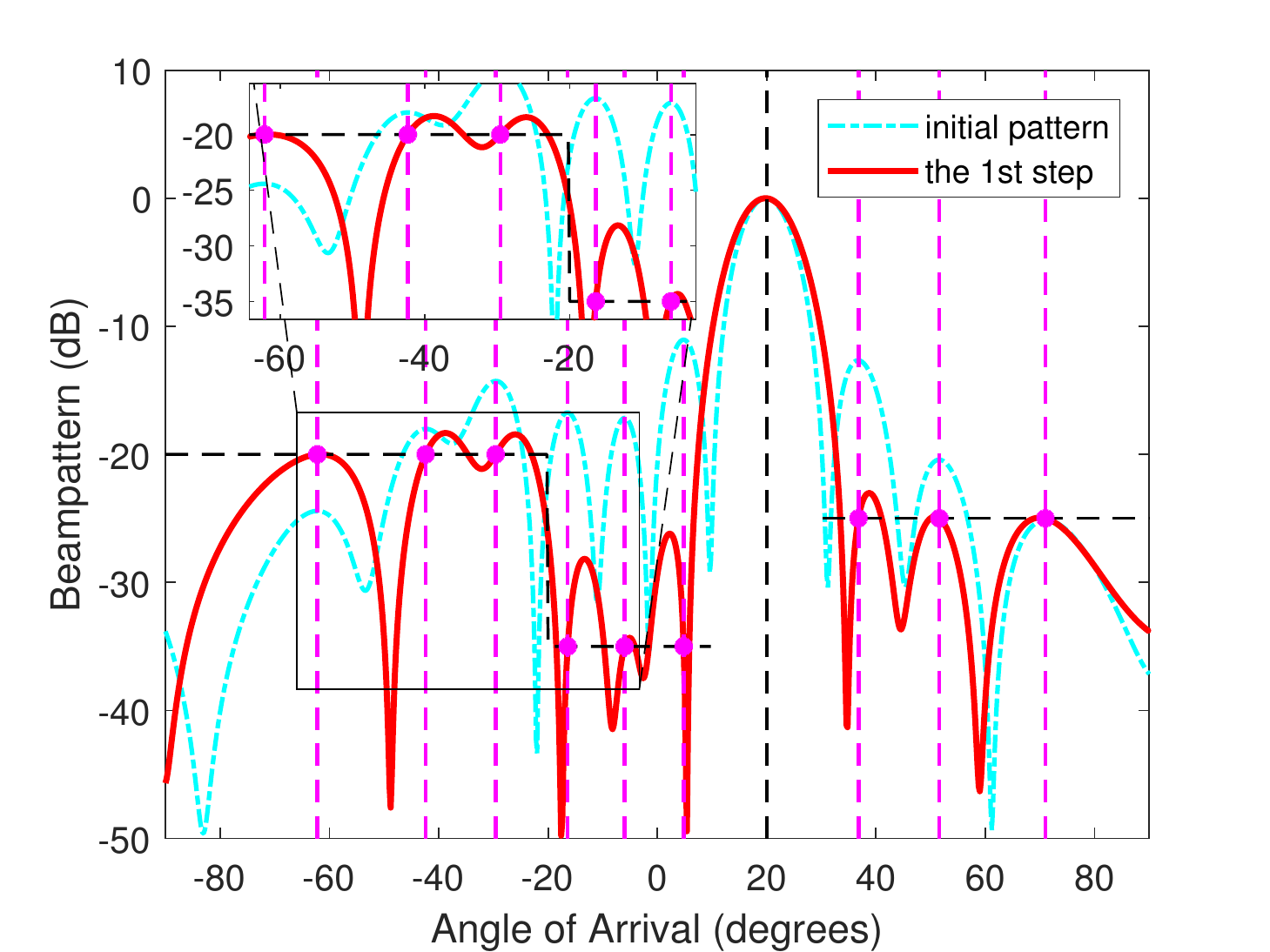}%
		\label{step1}}
	\hfil
	\subfloat[Synthesized pattern at the 2nd step]
	{\includegraphics[width=2.35in]{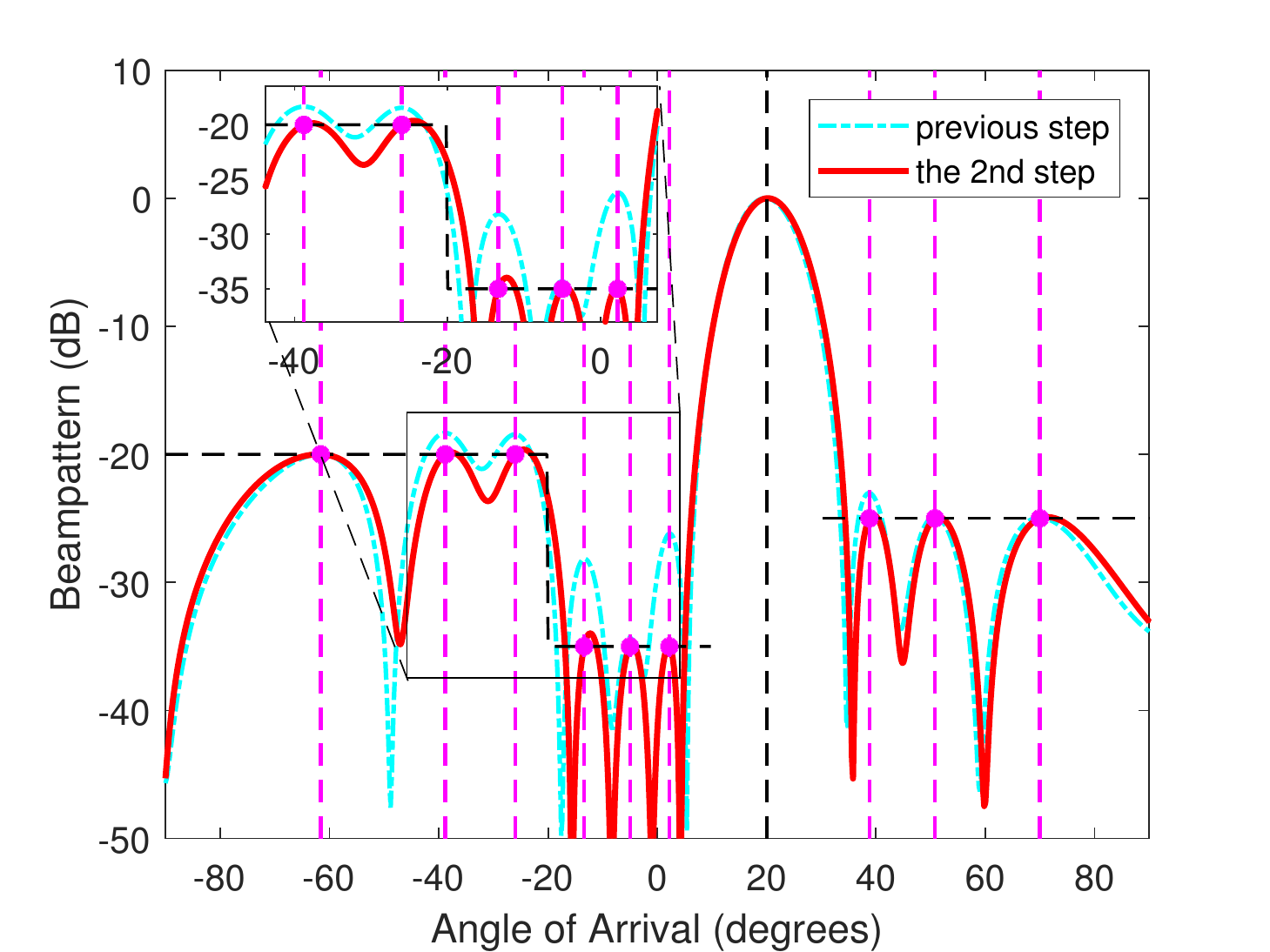}%
		\label{step2}}	
	\subfloat[Synthesized pattern at the 3rd step]
	{\includegraphics[width=2.35in]{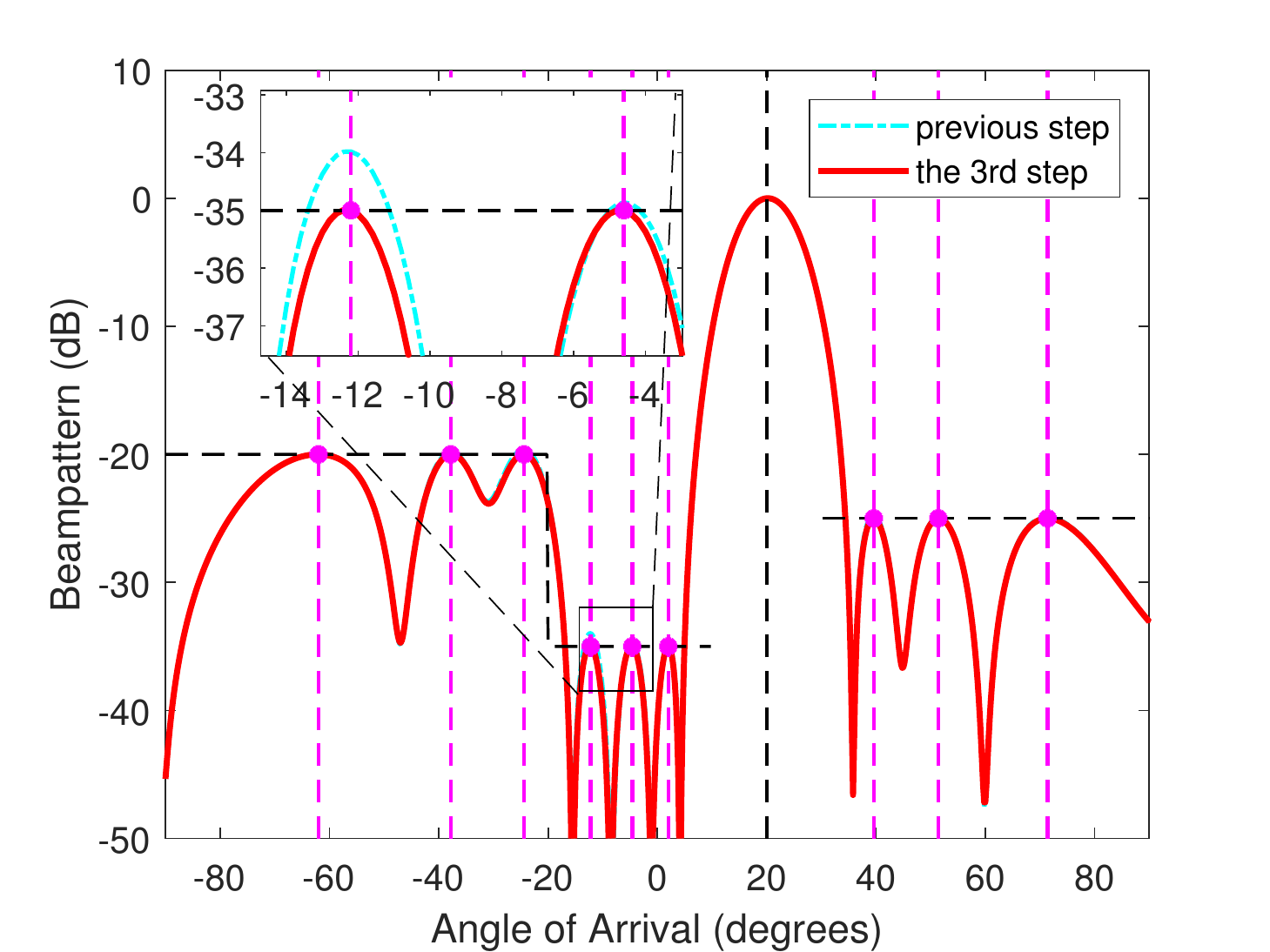}%
		\label{step3}}\\							
	\caption{Resultant patterns at different steps when carrying out a nonuniform sidelobe synthesis for a nonuniform linear array.}
	\label{dq}
\end{figure*}

\begin{figure}[!t]
	\centering
	\includegraphics[width=3.13in]{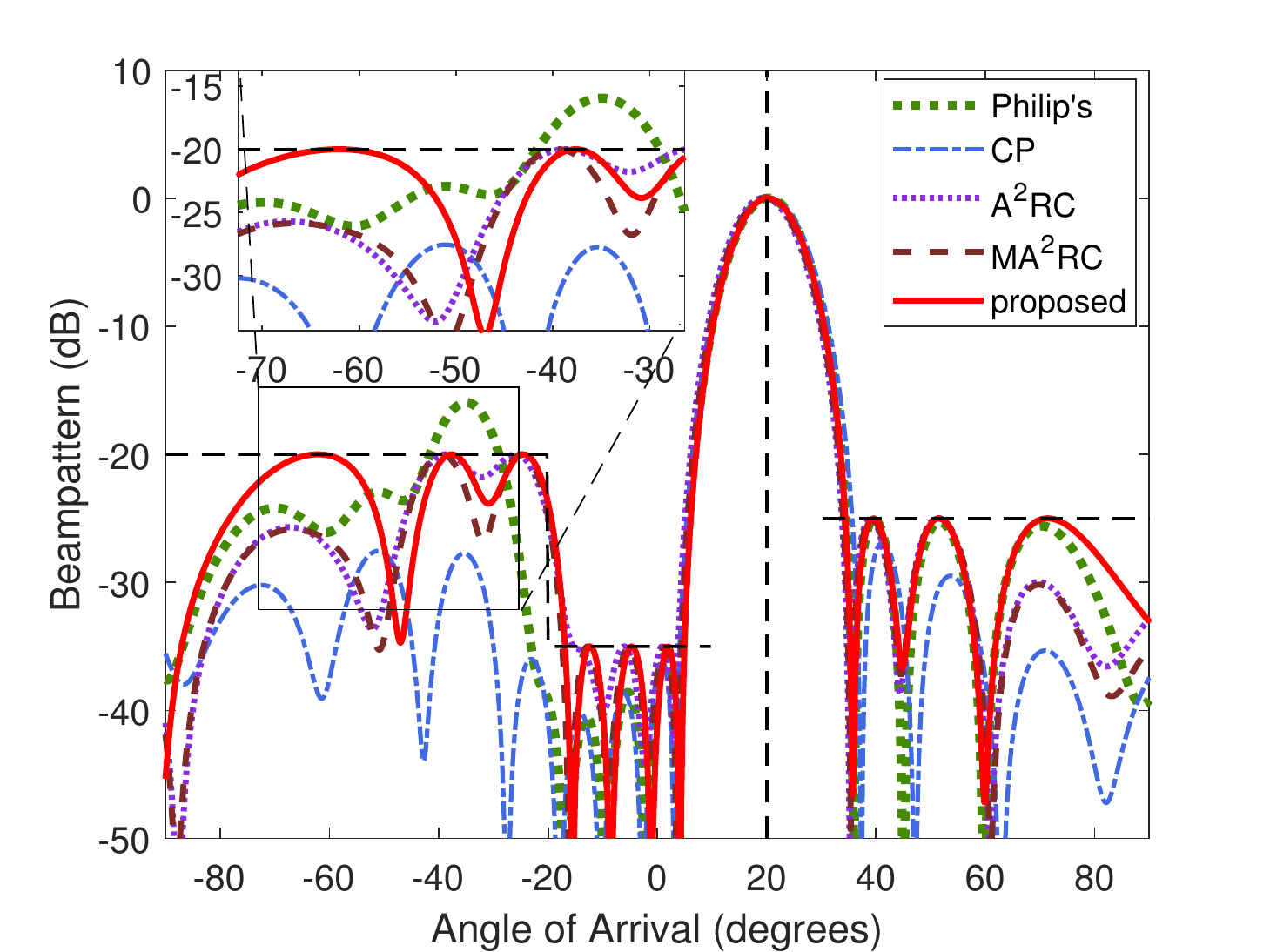}
	\caption{Resultant pattern comparison.}
	\label{Vcurve}
\end{figure}

More specifically, we set $ \theta_{1,1}=-45^{\circ} $, $ {\rho}_{1,1}=-40{\rm dB} $,
$ \theta_{1,2}=-5^{\circ} $ and $ {\rho}_{1,2}=-30{\rm dB} $ for the first step of
the response control.
Note that the same settings have been adopted in Section V.A in Part I \cite{p1},
where the single-point response control is realized in sequence. In this part, we
first conduct multi-point OPARC algorithm by using the iterative method described in Algorithm
\ref{cod1ed}. 
In the first iteration,
	the OPARC algorithm in \cite{p1} is applied to control the
	responses of $ \theta_{1,m} $ to their desired levels $ \rho_{1,m} $, $ m=1,2 $, one-by-one
	on $ m $.
	We have	
	$ {\beta}_{1,1,\star}=1.5683 $, $ {\beta}_{1,2,\star}=0.2504 $, which is
	the same as the results obtained in Section V.A in Part I \cite{p1}.
	Then, we continue our multi-point OPARC algorithm by conducting the above 
	iteration procedure for a number of times.
The curve of $ \beta_{\rm MAX} $ versus the iteration number
is depicted in Fig. \ref{beta}.
Note that the parameter $ \beta_{\rm MAX} $ measures
the maximal magnitudes of INRs of the newly assigned virtual interferences in the 
current iteration, as shown
in the 8th line of Algorithm \ref{cod1ed}.
From Fig. \ref{beta}, one can see that 
$ \beta_{\rm MAX} $ decreases with iteration.
Moreover, observation shows that it only requires 
five iterations to converge, i.e., $ {\beta}_{\rm MAX}\leq{\beta}_{\epsilon} $,
and the result is
$ \bar{\beta}_{1,1,\star}=1.4700 $ and $ \bar{\beta}_{1,2,\star}=0.2506 $,
which is, respectively, close to $ {\beta}_{1,1,\star} $ and $ {\beta}_{1,2,\star} $.
Now we test the performance of the C-ADMM approach.
The obtained $ \delta_{\rm MAX} $ in \eqref{condi}
reduces with the iteration, i.e., the procedure described 
	in \eqref{key05}-\eqref{criter}, as shown in Fig. \ref{delta},
and $ {\delta}_{\rm MAX}\leq{\delta} $ is met after about
130 iterations.
We obtain
$ {\bf h}_{1,\star}=[-0.1458 - j0.0203, -0.0687 - j0.0397]^{\mT} $.
Not surprisingly, it can be checked that the results of the above two approaches correspond
to the same weight vector.
Hence, the same beampatterns are synthesized for these two approaches as
shown in Fig. \ref{p3control}, from which one can see that the responses
of the two adjusted angles have been precisely controlled to their desired values.
Interestingly, when testing the ${\textrm M}{\textrm A}^2{\textrm{RC}}$,
the resulting pattern is completely the same as that of the multi-point OPARC algorithm.
We believe that this occurs not accidentally but with a reason that is, unfortunately, not clear yet.

\begin{table}[!t]
	\renewcommand{\arraystretch}{1.3}
	\caption{Obtained Parameter Comparison}
	\label{table5}
	\centering
	\begin{tabular}{c | c | c }
		\hline
		&${\textrm M}{\textrm A}^2{\textrm{RC}}$&Multi-point$~{\textrm{OPARC}}$ \\
		\hline
		$ D_1 ({\rm dB}) $ & $ 21.3110 $ &$ 5.7620 $ \\
		\hline
		$ D_2 ({\rm dB}) $ & $ 13.5149 $ &$ 10.0816 $ \\		
		\hline		
		$ J $ &$ 0.3132 $ &$ 0.1909 $\\
		\hline
		$ G_{1} ({\rm dB}) $&$ 10.0078 $  &$ 10.0078 $ \\
		\hline
		$ G_{2} ({\rm dB}) $ &$ 9.9192 $&$ 11.2550 $\\
		\hline		
	\end{tabular}
\end{table}

In the second step of the response control, we take 
$ \theta_{2,1}=7^{\circ} $, $ {\rho}_{2,1}=-25{\rm dB} $,
$ \theta_{2,2}=28^{\circ} $ and $ {\rho}_{2,2}=0{\rm dB} $.
When conducting the multi-point OPARC algorithm, we obtain 
$ \bar{\beta}_{2,1,\star}=0.2555 $ and $ \bar{\beta}_{2,2,\star}=-0.0804 $
for the iterative approach,
and find
$ {\bf h}_{2,\star}=[-0.1803 - j0.0653, -0.5434 - j0.9252]^{\mT} $
after implementing the C-ADMM method.
Again, the above two sets of results correspond to the same beampattern as shown
in Fig. \ref{p3control2}, where the resulting pattern of ${\textrm M}{\textrm A}^2{\textrm{RC}}$
is also displayed.
From Fig. \ref{p3control2}, one can see that all the adjusted angles
have been accurately controlled as expected, for the three approaches.
However, the mainlobe of the ultimate pattern of
	${\textrm M}{\textrm A}^2{\textrm{RC}}$ is distorted and
a high sidelobe level is resulted.
For comparison purpose, we have listed
several parameter measurements in Table \ref{table5},
from which one can see that the ${\textrm M}{\textrm A}^2{\textrm{RC}}$
method brings large values on both $ D_k $ ($ k=1,2 $) and $ J $, and
results a less array gain compared to the proposed multi-point OPARC algorithm.

\subsection{Array Pattern Synthesis Using Multi-point OPARC}
Starting from this subsection, the applications of multi-point OPARC
are simulated and the iterative approach in Section II. C
is adopted to illustrate the results. 
In this subsection, we focus upon the application of multi-point OPARC
to array pattern synthesis and give two representative examples for demonstration.

\subsubsection{Nonuniform Sidelobe Synthesis}
In the first example, 
the desired pattern has nonuniform sidelobes.
Fig. \ref{dq} shows the synthesized patterns of the proposed algorithm at different steps.
Clearly, in each synthesis step,
all the sidelobe peaks, i.e., $ {\Omega}_k $ in \eqref{edd},
are first determined from
the previously synthesized pattern. Notice that the response level
of a selected sidelobe peak can be either higher or lower (see Fig. \ref{step1}
for reference) than its desired level.
It has been shown in Fig. \ref{dq} that it only requires
3 steps, i.e., $ k=3 $, to synthesize a satisfactory beampattern.

\begin{figure*}[!t]
	\centering
	\subfloat[Synthesized pattern at the 1st step]
	{\includegraphics[width=2.35in]{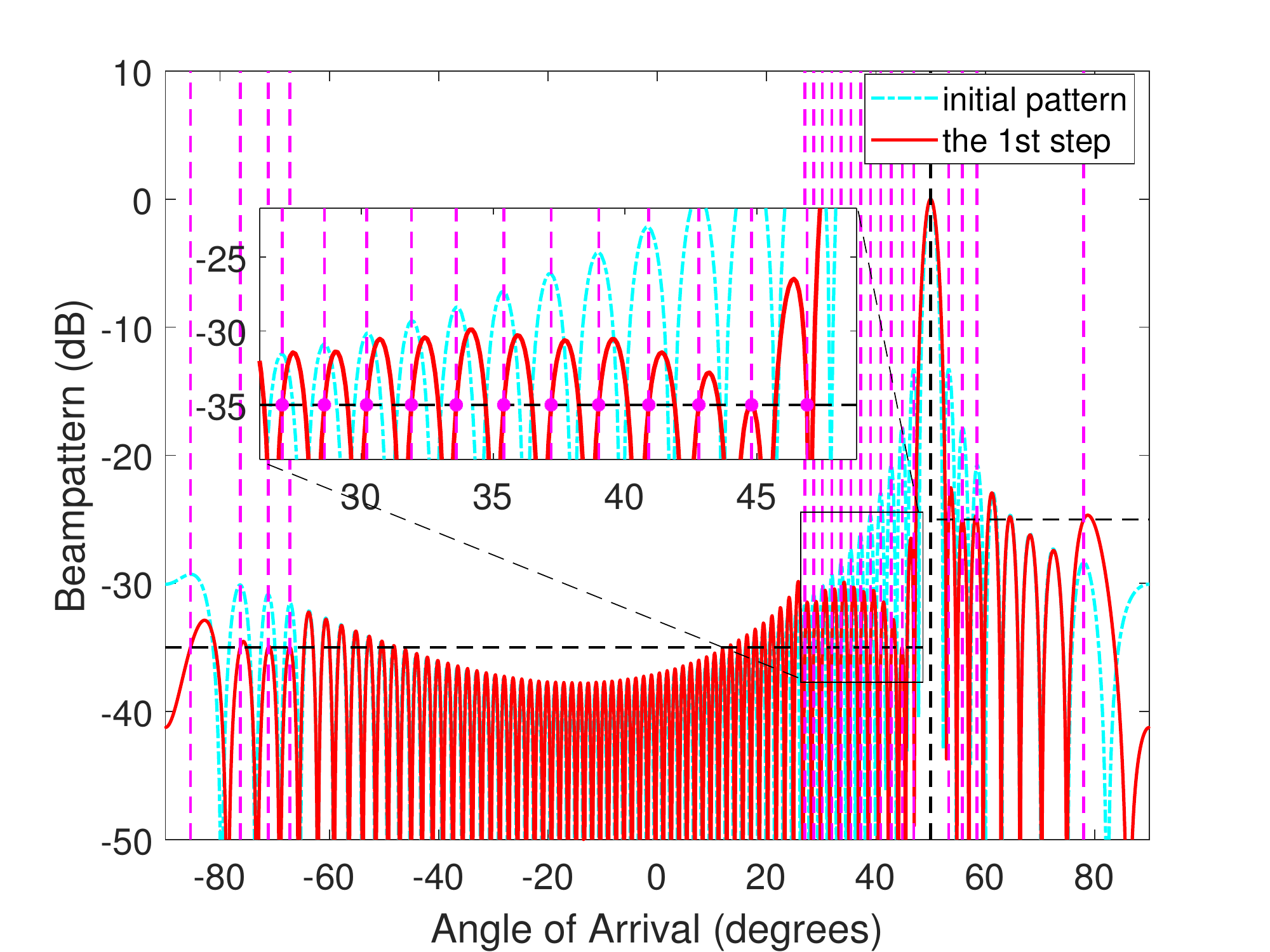}%
		\label{ULAstep1}}
	\hfil
	\subfloat[Synthesized pattern at the 2nd step]
	{\includegraphics[width=2.35in]{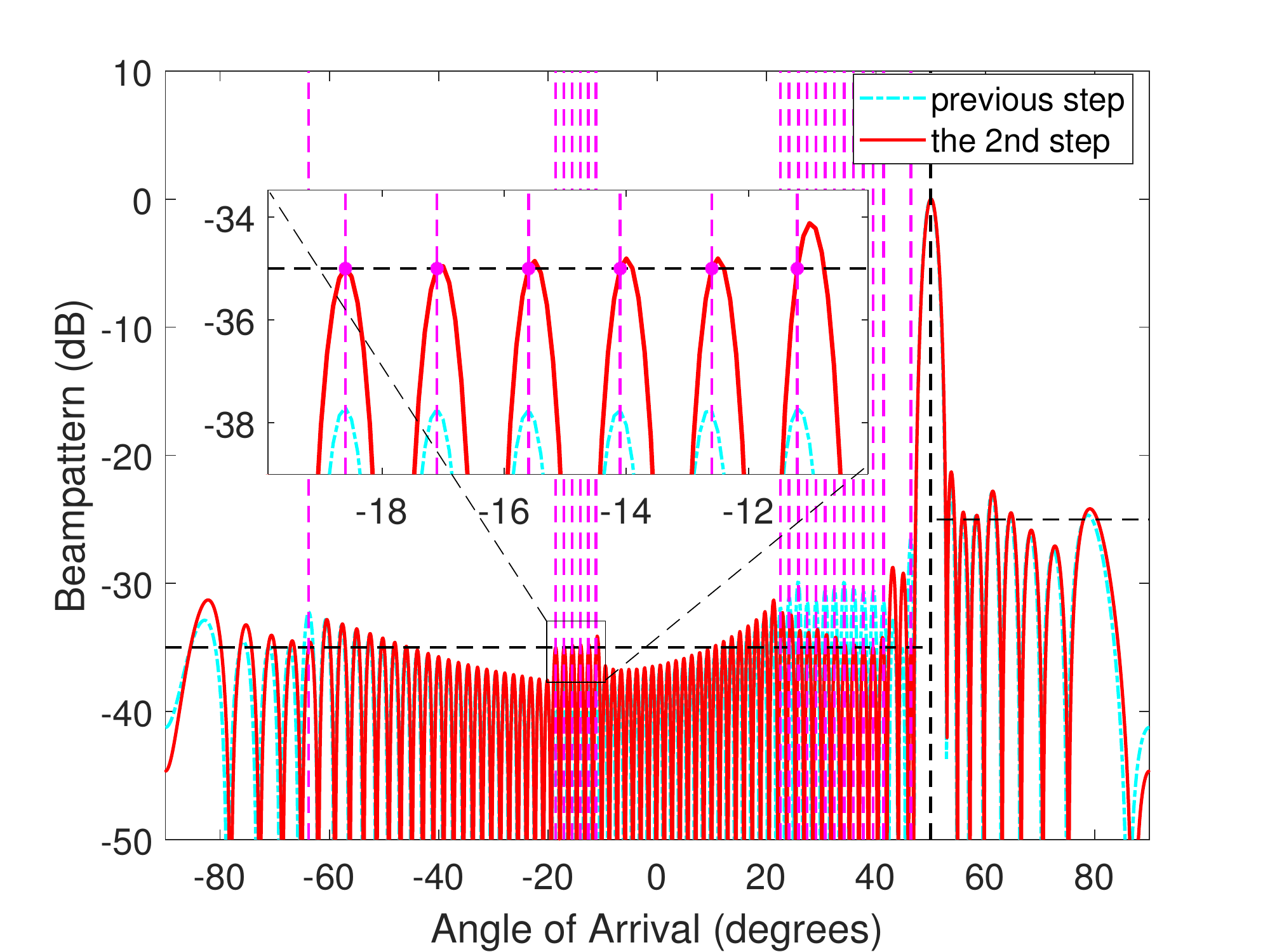}%
		\label{ULAstep2}}
	\hfil
	\subfloat[Synthesized pattern at the 11th step]
	{\includegraphics[width=2.35in]{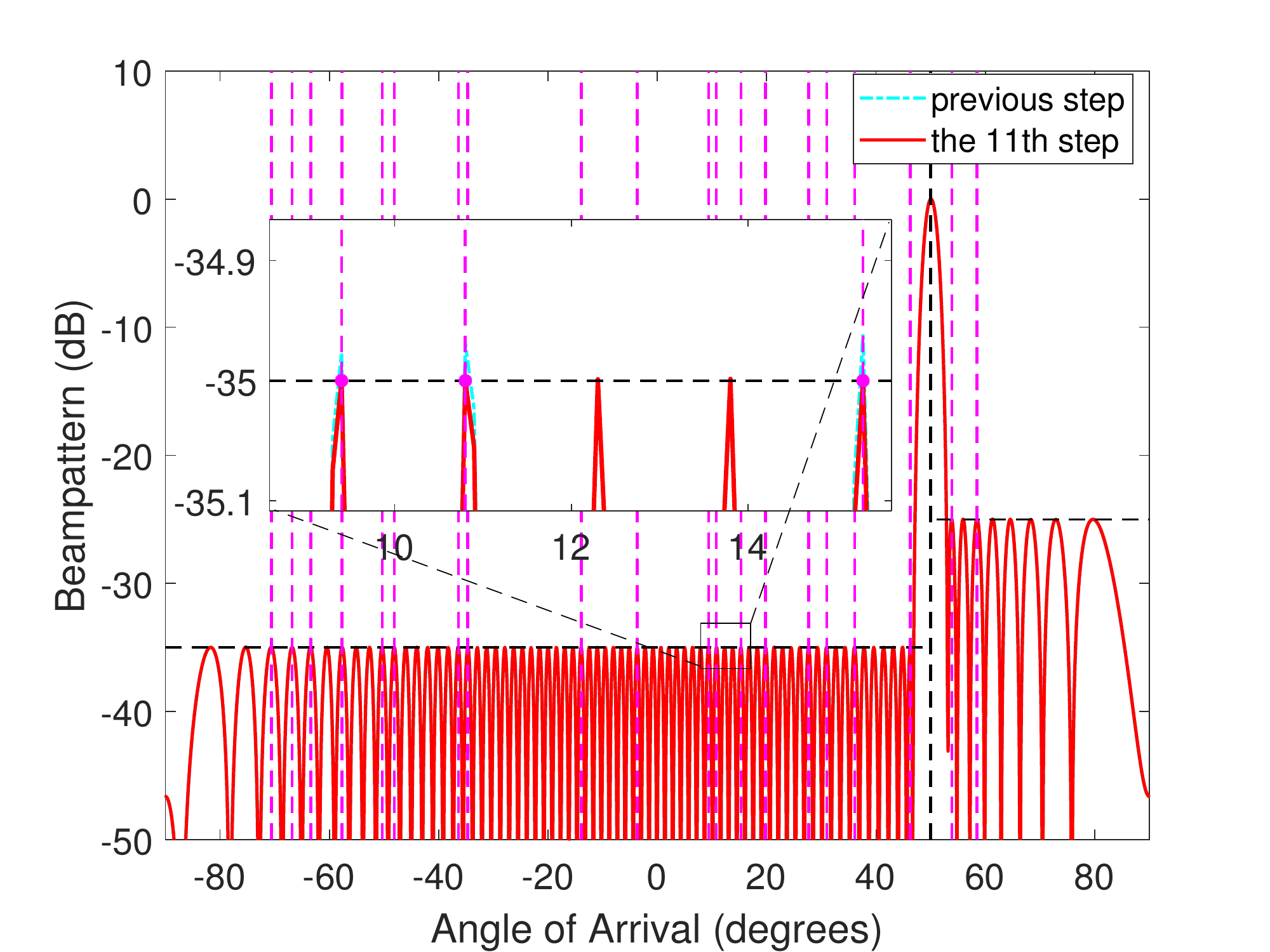}%
		\label{ULAstep11}}\\							
	\caption{Resultant patterns at different steps when carrying out a nonuniform sidelobe synthesis for a large uniform linear array.}
	\label{dq2}
\end{figure*}


For comparison, the resulting patterns of the proposed algorithm,
Philip's method in \cite{snrf12}, convex programming (CP) method in \cite{snrf13},
$ {\textrm A}^2\textrm{RC} $ method (after carrying out 30 steps) in \cite{snrf41}
and $ {\textrm M}{\textrm A}^2\textrm{RC} $ method (after carrying out 3 steps) 
in \cite{ref201} are displayed in Fig. \ref{Vcurve}.
As expected, we can see that the pattern envelopes of Philip¡'s method and CP
method are not aligned with the desired level, since they cannot
control the beampattern precisely according to the required
specifications.
Although $ {\textrm A}^2\textrm{RC} $ and $ {\textrm M}{\textrm A}^2\textrm{RC} $
have the ability to precisely control the given array responses,
the obtained sidelobe peaks are not aligned with the desired ones either,
since only the sidelobe peaks higher than the desired levels 
are
selected and adjusted in these two approaches.

\begin{table}[!t]
	\renewcommand{\arraystretch}{1.3}
	\caption{Execution Time Comparison When Conducting a Large-array Pattern Synthesis}
	\label{tablenonuniform0}
	\centering
	\begin{tabular}{c | c | c | c | c | c}
		\hline
		& Philip's &$ ~~~ $CP$ ~~~ $ &	
		$ ~{\textrm {A}}^2\textrm{RC}~ $ & $ {\textrm {M}}{\textrm {A}}^2\textrm{RC} $ &proposed  \\
		\hline	
		$ T(\rm sec) $& 2.22	 &12.36	 &	3.55 & 2.55 & 0.05					  \\
		\hline
	\end{tabular}
\end{table}

\subsubsection{Large Array Consideration}
In this example, pattern synthesis for
a large linearly half-wavelength-spaced array with $ N=80 $ isotropic elements is considered.
The desired pattern steers at $ \theta_0=50^{\circ} $ with
nonuniform sidelobes. More specifically, the upper level is $ -35{\rm dB} $ in the sidelobe region $ [-90^{\circ},50^{\circ}) $ and $ -25{\rm dB} $ in the rest of the sidelobe region.

Fig. \ref{dq2} demonstrates several intermediate results of the proposed algorithm.
In every step, we select $ C_k=20 $ sidelobe peak angles
(see Eqn. \eqref{theta} and \eqref{theta1} for details) and then adjust
their responses to the desired levels by using multi-point OPARC algorithm.
Simulation result shows that it only requires $ 11 $ steps, i.e., $ k=11 $,
to synthesize
a qualified pattern, see the ultimate pattern in Fig. \ref{ULAstep11}
for reference. The execution times of various methods are 
provided in Table \ref{tablenonuniform0},
where the superiority of the proposed algorithm can be clearly observed.


\subsection{Multi-constraint Adaptive Beamforming Using Multi-point OPARC}
In this subsection, the multi-constraint adaptive beamforming is 
realized by using the multi-point OPARC algorithm.
For simplicity,
a perfect knowledge of the data covariance matrix is assumed.

\begin{figure}[!tpb]
	\centering\subfloat[The first case]
	{\includegraphics[width=3.13in]{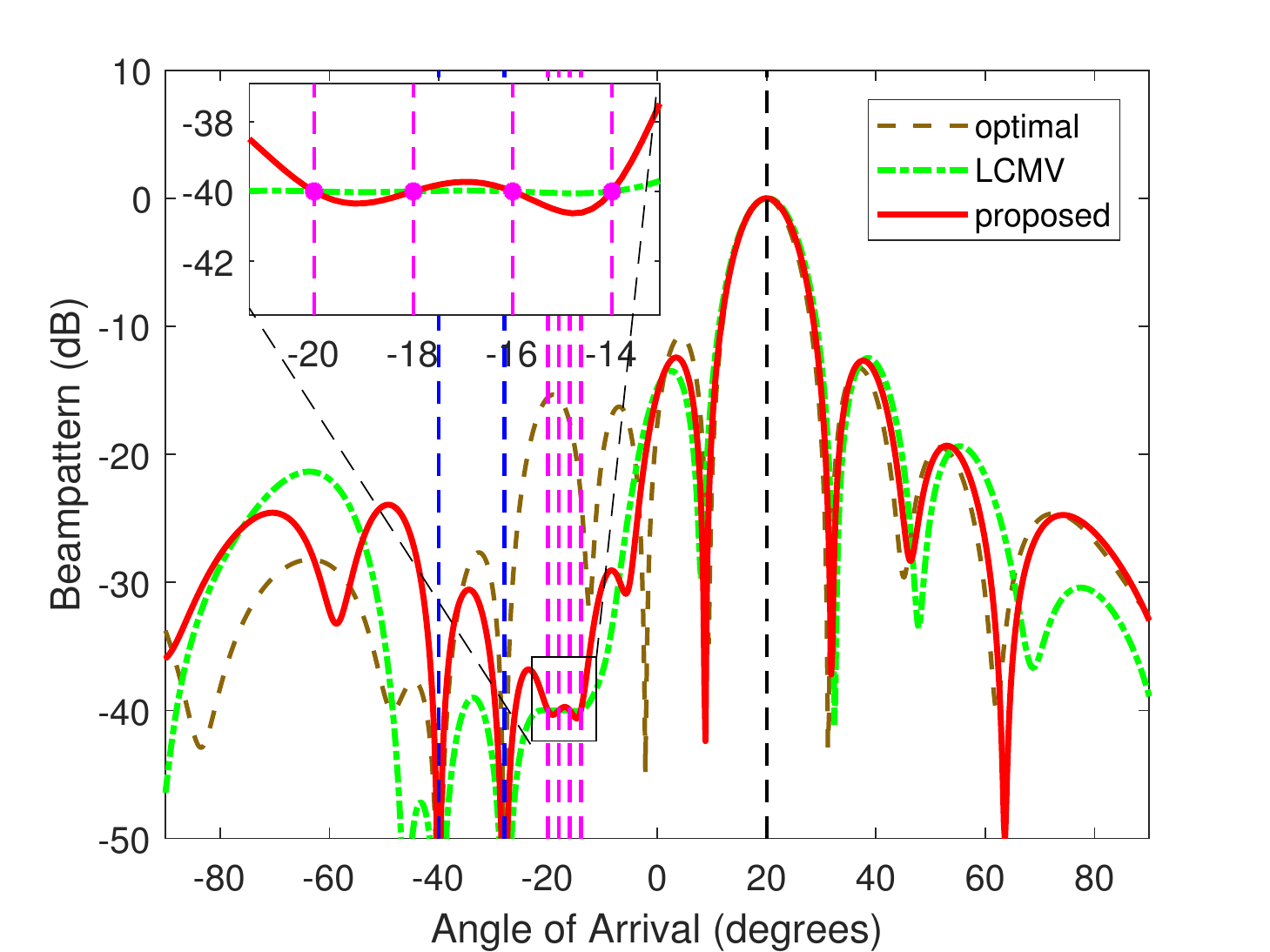}%
		\label{LCMV1}}\
	\centering\subfloat[The second case]
	{\includegraphics[width=3.13in]{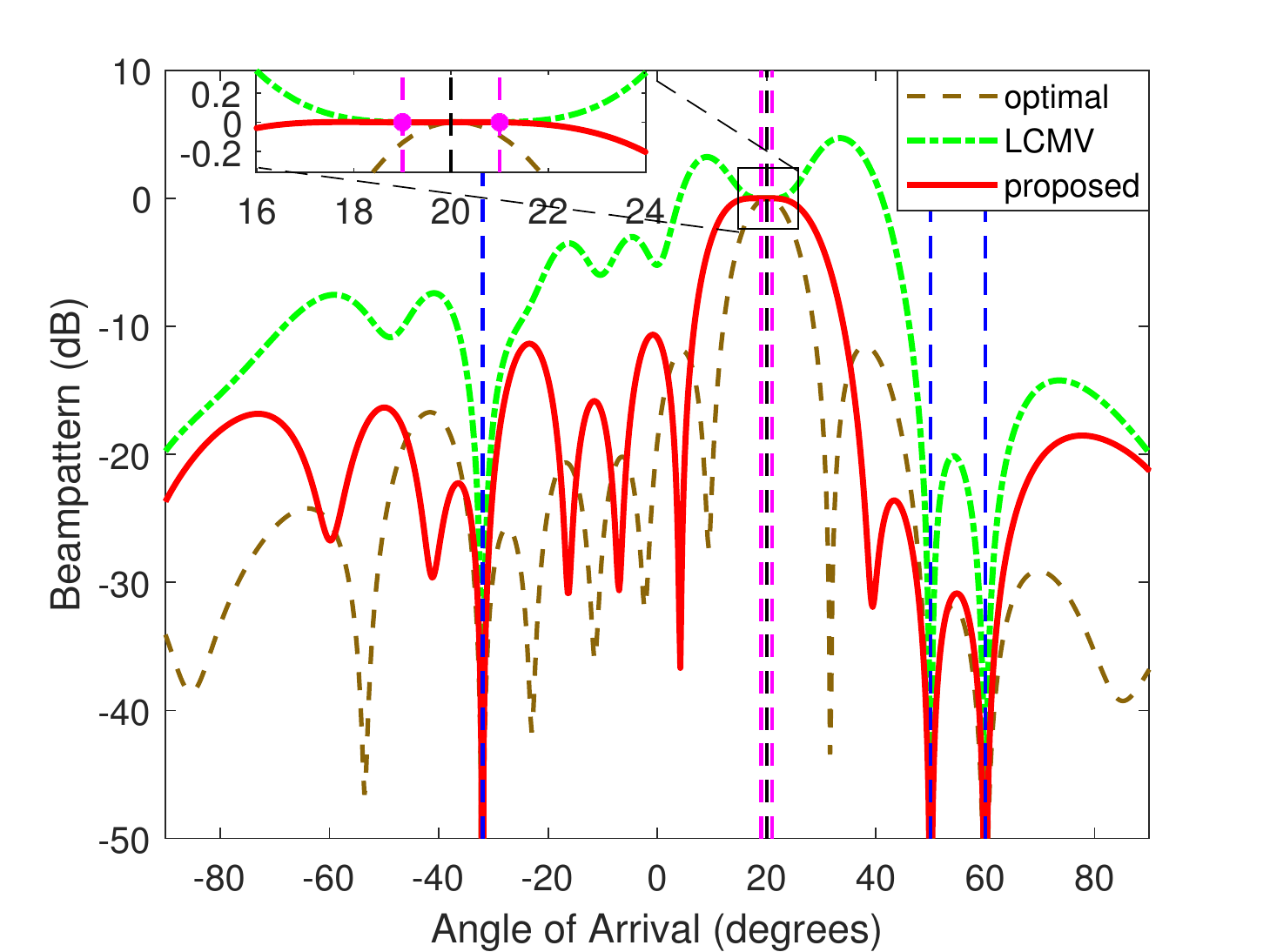}%
		\label{LCMV2}}			
	\caption{Result comparison of multi-constraint adaptive beamforming.}
	\label{dq1d}
\end{figure}

\subsubsection{Sidelobe Constraint}
In the first case, four sidelobe constraints are required.
More specifically, the response levels of $ -20^{\circ} $, $ -18^{\circ} $,
$ -16^{\circ} $ and $ -14^{\circ} $ are expected to be all $ -40{\rm dB} $.
Two interferences are impinged from $ -40^{\circ} $ and $ -28^{\circ} $ with
INRs $ 30{\rm dB} $ and $ 25{\rm dB} $, respectively.

Fig. \ref{LCMV1} displays the results of the optimal beamformer with no sidelobe constraint,
the LCMV method \cite{lcmv} and the proposed one.
Clearly, both the LCMV beamformer and the proposed algorithm
are able to shape deep nulls at the directions of interferences (see the blue line).
Meanwhile, the given sidelobe constraints are well satisfied for both.
When considering the output SINR, we have $ {\rm SINR}=19.5601{\rm dB} $ for
the LCMV method and $ {\rm SINR}=19.6906{\rm dB} $
for the proposed one.
We can see that the proposed beamformer brings an improvement on the output SINR compared to
the LCMV beamformer.

\subsubsection{Mainlobe Constraint}
In the second case, two constraints are imposed in the mainlobe region.
The constraint angles are $ 19^{\circ} $ and $ 21^{\circ} $,
and both of the desired levels are $ 0{\rm dB} $.
There are three interferences coming from $ -32^{\circ} $, $ 50^{\circ} $ and $ 60^{\circ} $ with
an identical INR $ 30{\rm dB} $.

Fig. \ref{LCMV2} depicts the resultant patterns.
One can see that the obtained pattern of the LCMV method is severely distorted,
although the two prescribed constraints are satisfied and the three interferences
are rejected. The corresponding output SINR is $ 11.1767{\rm dB} $.
Observing the resulting pattern of the proposed algorithm, the two-point constraint
is well satisfied and a flat-top mainlobe is shaped with no distortion occurred.
The corresponding output SINR is $ 17.1260{\rm dB} $, which is much higher than that of the 
LCMV method.

\begin{figure}[!tpb]
	\centering\subfloat[The first case]
	{\includegraphics[width=3.13in]{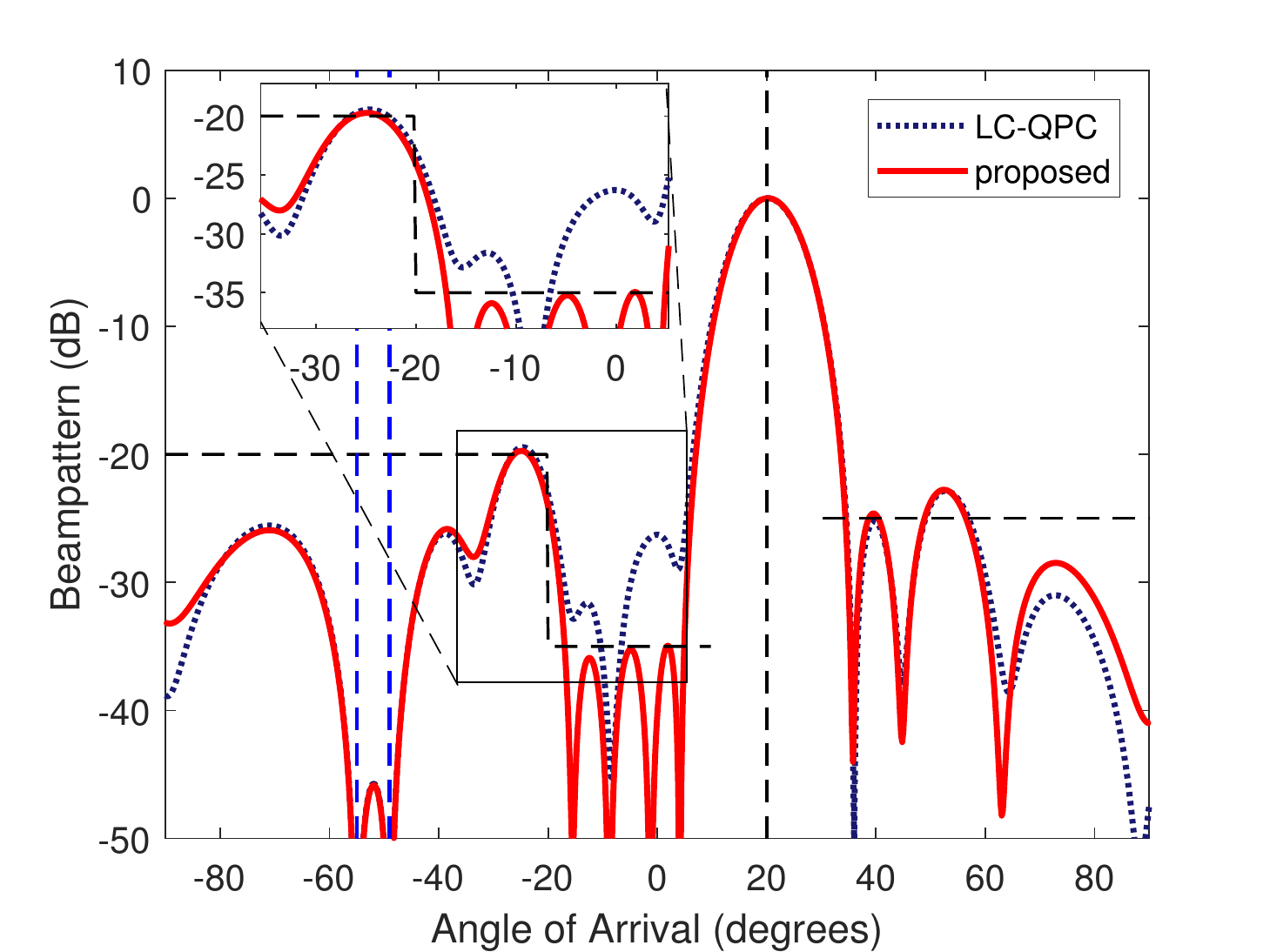}%
		\label{QPC}}\
	\centering\subfloat[The second case]
	{\includegraphics[width=3.13in]{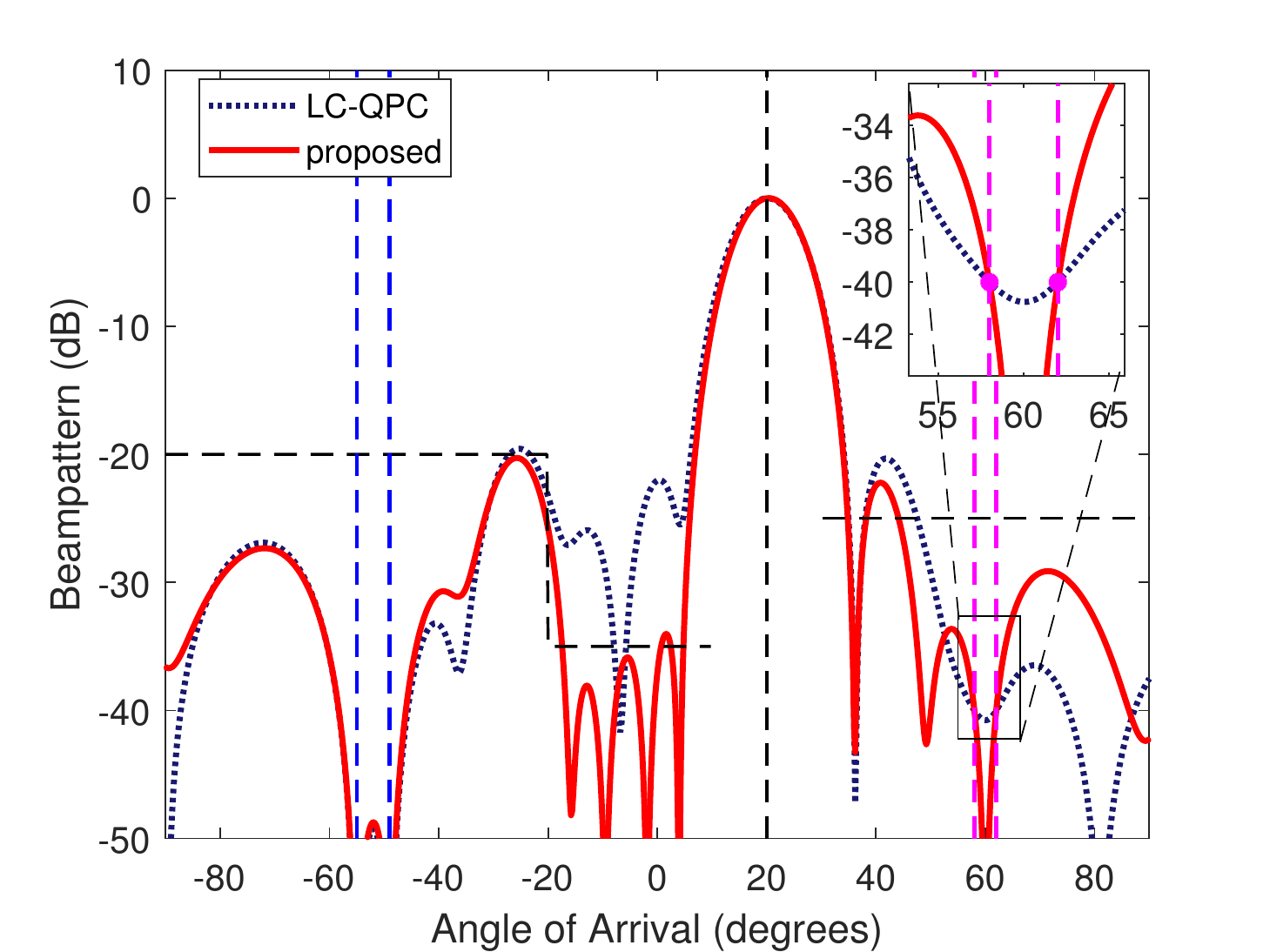}%
		\label{QPCconstraint}}			
	\caption{Result comparison of quiescent pattern control.}
	\label{dq1dd}
\end{figure}

\subsection{Quiescent Pattern Control Using Multi-point OPARC}
In this subsection, we test
the performance of the multi-point OPARC based quiescent pattern control algorithm.
The desired quiescent
pattern has a nonuniform sidelobe level as depicted with black dash lines
in Fig. \ref{Vcurve}.

In our proposed algorithm, quiescent pattern synthesis and quiescent pattern control are
jointly designed by the multi-point OPARC algorithm.
We have detailed the off-line synthesis procedure in Section IV.B
and illustrated the obtained quiescent pattern by red line in Fig. \ref{Vcurve}.
Suppose that two interferences come
	from $ -55^{\circ} $ and $ -49^{\circ} $ with INRs $ 30{\rm dB} $.
The obtained adaptive response pattern is shown in Fig. \ref{QPC},
where we can observe that two nulls are formed at the
directions of the real interferences, and the resultant sidelobe is
close to the quiescent one.
The obtained output SINR is $ 19.2984{\rm dB} $ for the proposed algorithm.

For comparison purpose, the classical linearly-constraint based quiescent pattern control
approach (denoted as LC-QPC method for briefness) in \cite{con6} is also demonstrated,
by using the same synthesized quiescent pattern in Fig. \ref{Vcurve}.
The resulting pattern of LC-QPC is displayed in
Fig. \ref{QPC}, where we find that an obvious perturbation is caused in the sector $ [-15^{\circ},0^{\circ}] $ and the overall shape can not be well maintained compared to the desired one.
The obtained output SINR is $ 19.2161{\rm dB} $, which is lower than
that of the proposed algorithm.

Now we take extra fixed constraints into consideration by
	restricting the response levels at directions 
	$ 58^{\circ} $ and $ 62^{\circ} $ to be all $ -40{\rm dB} $.
The results of the proposed algorithm and the LC-QPC method are presented in
Fig. \ref{QPCconstraint}, where
we observe that both of these two methods are able to reject the
undesirable interferences with the prescribed constraints being satisfied.
The same as before, the proposed algorithm maintains a more desirable shape than that of
the LC-QPC method.
When taking the output SINR into account, the corresponding values are, respectively,
$ 19.2382{\rm dB} $ (for the proposed algorithm) and $ 19.0967{\rm dB} $ (for
the LC-QPC method).
The advantage of the proposed algorithm is verified again.

\section{Conclusions}
In this paper, the optimal and precise array
response control (OPARC) algorithm proposed in Part I \cite{p1} has been extended
from a single point per step to a multi-points per step.
Two computationally attractive multi-point OPARC algorithms
have been proposed, by which the responses of multiple angles can be adjusted.
In addition, several
applications of the multi-point OPARC algorithm to array signal processing have been presented,
and an innovate concept of normalized covariance matrix loading (NCL) has been
developed.
Simulation results have been provided to validate the effectiveness and superiority of the proposed algorithms under different situations.

\vspace*{-0.326\baselineskip}
\bibliography{FinalVersionPart1ofOPARC20171225}
\end{document}